\documentclass[aps,prx,reprint,superscriptaddress,floatfix]{revtex4-2}

\usepackage[utf8]{inputenc}
\usepackage{mathtools}
\usepackage{physics}
\usepackage{hyperref}
\usepackage{amsmath,amsthm,amssymb,amsfonts}
\usepackage[utf8]{inputenc}
\usepackage{mathrsfs}
\usepackage{verbatim}
\usepackage{graphicx,graphics}
\usepackage{svg}
\usepackage{subfigure}
\usepackage{wrapfig}
\usepackage{hyperref}
\usepackage{gensymb}
\usepackage{dsfont}
\usepackage{upgreek}
\usepackage{braket}
\usepackage[useregional]{datetime2}
\usepackage[symbol]{footmisc}
\usepackage{titlesec}

\hypersetup{
  colorlinks=true,
  urlcolor=blue,
  citecolor=blue,
}
\usepackage[margin=1.0in]{geometry}
\usepackage{wrapfig}
\usepackage{tabularx}
\usepackage{setspace}

\begin{document}

\preprint{}

\title{Scalable parallel measurement of individual nitrogen-vacancy centers}

\author{Matthew Cambria}
\thanks{These authors contributed equally to this work.}
\author{Saroj Chand}
\thanks{These authors contributed equally to this work.}
\author{Caitlin Reiter}
\affiliation{Department of Physics, University of California, Berkeley, CA, USA}
\author{Shimon Kolkowitz}
\email{kolkowitz@berkeley.edu}
\affiliation{Department of Physics, University of California, Berkeley, CA, USA}
\affiliation{Department of Physics, University of Wisconsin, Madison, WI, USA}

\date{\today}

\begin{abstract}
The nitrogen-vacancy (NV) center in diamond is a solid-state spin defect that has been widely adopted for quantum sensing and quantum information processing applications. Typically, experiments are performed either with a single isolated NV center or with an unresolved ensemble of many NV centers, resulting in a trade-off between measurement speed and spatial resolution or control over individual defects. In this work, we introduce an experimental platform that bypasses this trade-off by addressing multiple optically resolved NV centers in parallel. We perform charge- and spin-state manipulations selectively on multiple NV centers from within a larger set, and we manipulate and measure the electronic spin states of over 100 NV centers in parallel. We show that the high signal-to-noise ratio of the measurements enables the detection of shot-to-shot pairwise correlations between the spin states of 108 NV centers, corresponding to the simultaneous measurement of 5,778 unique correlation coefficients. We discuss how our platform can be scaled to parallel experiments with thousands of individually resolved NV centers. These results enable parallelized high-throughput sensing experiments that retain the nanoscale spatial resolution of single defects, and will thereby help to unlock advances in applications such as single-molecule NMR and characterization of integrated circuits. In addition, our approach to multiplexing provides a natural platform for the application of recently developed correlated sensing techniques.
\end{abstract}

\maketitle

The nitrogen-vacancy (NV) center in diamond has emerged as a solid-state platform with applications in quantum sensing \cite{maze2008nanoscale, dolde2011electric, rondin2012nanoscale, levine2019principles, segawa2023nanoscale} and quantum information processing \cite{dutt2007quantum, bernien2013heralded, abobeih2022fault, fung2024toward}.
The NV center is notable for its robust and optically accessible electronic spin ground-state, which can exhibit coherence times of several milliseconds at room temperature \cite{balasubramanian2009ultralong, herbschleb2019ultra}. As a result, the NV center is now employed as a versatile quantum sensor with applications in condensed matter physics \cite{vool2021imaging, huang2023revealing, bhattacharyya2024imaging}, biology \cite{choi2020probing, nie2021quantum}, chemistry \cite{abendroth2022single, liu2022surface}, and electronics \cite{turner2020magnetic, garsi2024three}. Almost all experiments with NV centers are performed either by interrogating one spatially resolved NV center at a time or by interrogating ensembles of many unresolved NV centers.
Measurements with single NV centers offer nanoscale spatial resolution, 
but are typically slow, as each experiment only returns up to one bit of information, and measurements at different locations must be conducted in series in imaging applications. In addition, measurements with single NV centers only provide information about one location at a time, and therefore cannot access spatial correlations in fluctuating fields \cite{ariyaratne2018nanoscale, finco2021imaging}. 
In contrast, measurements with ensembles of many NV centers can be comparatively fast, but their spatial resolution is limited by optical diffraction, and information is lost in the process of averaging over the ensemble \cite{glenn2018high, ziffer2024quantum}. In particular, wide-field implementations in which a dense 2D layer of NV centers is imaged onto a camera have become an important tool in quantum sensing \cite{scholten2021widefield}, but are not suitable for applications that require readout and control at the level of individual NV centers. A recent work demonstrated a hybrid approach using programmable 532-nm illumination with a digital digital micromirror device (DMD) to conduct parallel spin experiments with multiple optically resolved NV centers imaged on a continuously exposed CMOS camera, at the cost of a reduction in the optically detected magnetic resonance contrast \cite{cai2023toward}. 

Several recent works have developed a new measurement modality in which shot-to-shot correlations between signals from several NV centers are used to obtain information about the environment which is otherwise difficult to access \cite{rovny2022nanoscale, ji2024correlated, delord2024correlated, huxter2024multiplexed}. These correlated sensing techniques require both the ability to work with multiple NV centers simultaneously, and the ability to associate an experimental signal with a specific NV center. 
In Ref.~\cite{rovny2022nanoscale}, correlations between the spin states of two optically resolved NV centers were measured using a confocal microscope modified with a 50:50 beamsplitter to allow for independent addressing and readout of each NV. Because half the light from each NV is lost and each NV requires its own lasers and photodetector, the scalability of this approach is limited. The techniques developed in Refs.~\cite{ji2024correlated, delord2024correlated, huxter2024multiplexed} are designed to measure correlations between NV centers within a single diffraction-limited spot, and require that the NV centers in the spot have frequency-resolved optical \cite{ji2024correlated, delord2024correlated} or microwave \cite{huxter2024multiplexed} transitions. 

\begin{figure*}[!t]
    \includegraphics[width=1\textwidth]{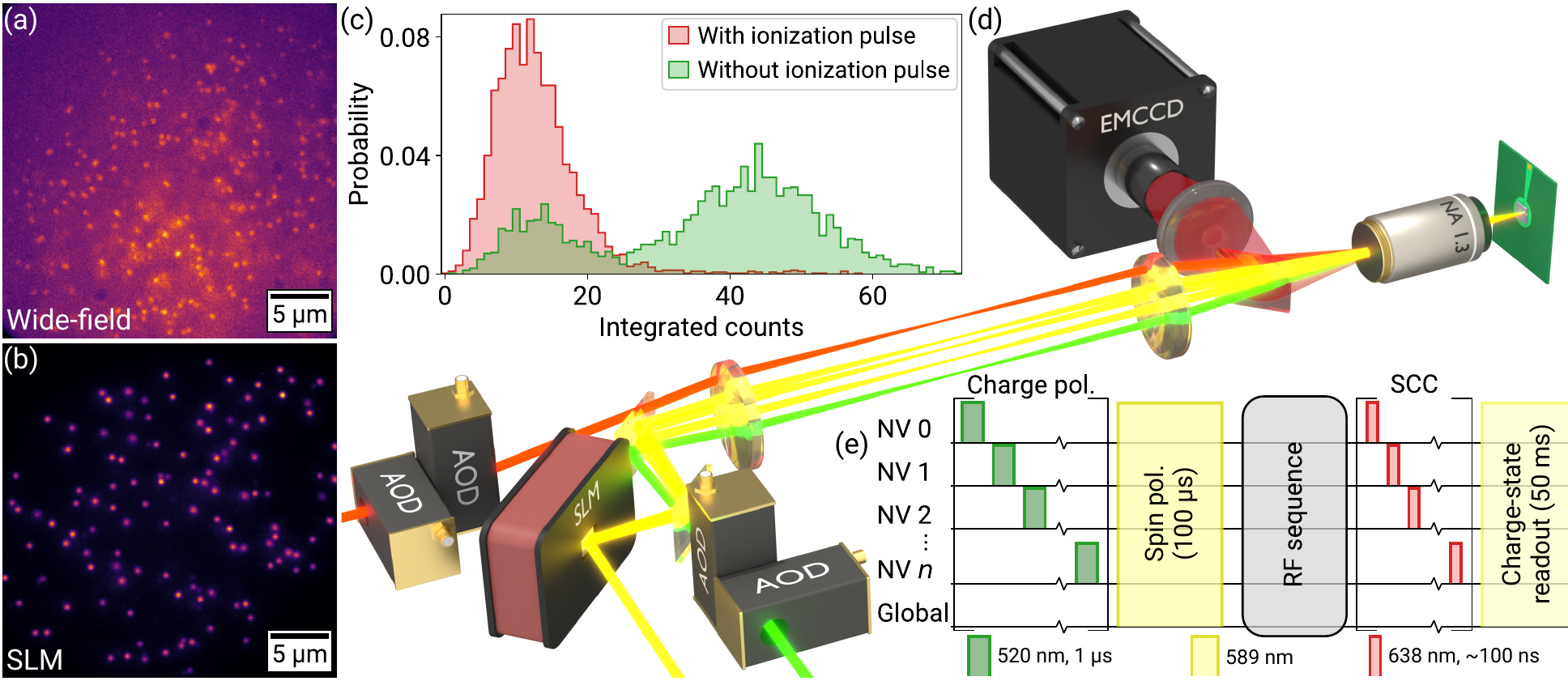}
    \caption{Experimental platform.
    (a) Composite wide-field image under 520-nm illumination showing the region of the diamond used in this work (Appendix~\ref{subsec:app-identification}).
    (b) Same region imaged under 589-nm illumination patterned with a spatial light modulator (SLM). 117 diffraction-limited target spots are illuminated, of which 108 are charge-stable single NV centers exhibiting consistent optically detected magnetic resonance signals. 
    (c) Charge-state manipulation and readout. 
    The histograms show the probability of recording a given number of photons from a single NV center in a 50 ms exposure after polarization into NV$^{-}$ (green) with 520 nm illumination or NV$^{0}$ (red) with 638 nm illumination. 
    (d) Illustration of the experimental apparatus. The SLM allows for an arbitrary set of NV centers from within the field-of-view to be addressed in parallel by splitting the incident 589-nm illumination into an arbitrary pattern of beams, each of which targets a single NV center. Switching the frequencies of the drive tones applied to the acousto-optic deflectors (AODs) rapidly repositions the 520-nm and 638-nm laser beams between target NV centers (within 10~{\textmu}s). A microwave antenna \cite{sasaki2016broadband} delivers global microwave pulses with Rabi frequencies of approximately 10 MHz. Fluorescence from the NV centers is imaged onto an EMCCD camera. 
    (e) Experimental sequence for parallel spin measurements. The green, yellow, and red blocks indicate illumination with 520 nm, 589 nm, and 638 nm respectively. The global microwave sequence (gray block) varies depending on the experiment. Charge-polarization and spin-to-charge conversion (SCC) steps are conducted by applying optical pulses to the \(n\) NV centers one at a time in series. Spin polarization is conducted as a separate step from charge polarization \cite{wirtitsch2023exploiting} and uses the same 589-nm illumination pattern as in charge-state readout up to a difference in global intensity. Spin polarization and charge-state readout are performed in parallel across all target NV centers. See Appendix~\ref{sec:app-exp_details} for further details of the apparatus and sequence. 
    }
    \label{fig:scheme}
\end{figure*}

Motivated by these considerations, in this work we present an experimental platform for the simultaneous parallel interrogation of many spatially resolved NV centers with high signal-to-noise ratios in a scalable manner. 
Our approach is inspired by the techniques underlying reconfigurable arrays of neutral atoms in optical tweezers, which have developed rapidly over the past decade \cite{schlosser2012fast, xia2015randomized, endres2016atom, barredo2016atom, kaufman2021quantum}. 
In particular, we use NV center-selective charge-state manipulations and single-shot charge-state readout recorded under patterned illumination onto an electron-multiplying CCD (EMCCD) camera to conduct parallel charge- and spin-state experiments with 108 NV centers. 
Further, we demonstrate the ability to measure pairwise correlations between all of the NV center spin states simultaneously, yielding 5,778 unique correlation coefficients. 
Our results demonstrate that nanoscale quantum sensing with single NV centers can be performed in a scalable manner. We anticipate that this will enable new applications in areas where data collection rates with single NV centers are prohibitively slow, such as biological systems with low signal levels \cite{zhang2021toward, du2024single} and in condensed matter samples 
which exhibit correlated dynamics or where high-throughput data collection is required.
We show that our platform can also be used for covariance sensing at scale, enabling experimentally practical measurements of systems with micron-scale correlation lengths \cite{casola2018probing, xu2023recent, rovny2024nanoscale, kolkowitz2015probing, jenkins2019single}.

\begin{figure*}[!t]
    \includegraphics[width=1\textwidth]{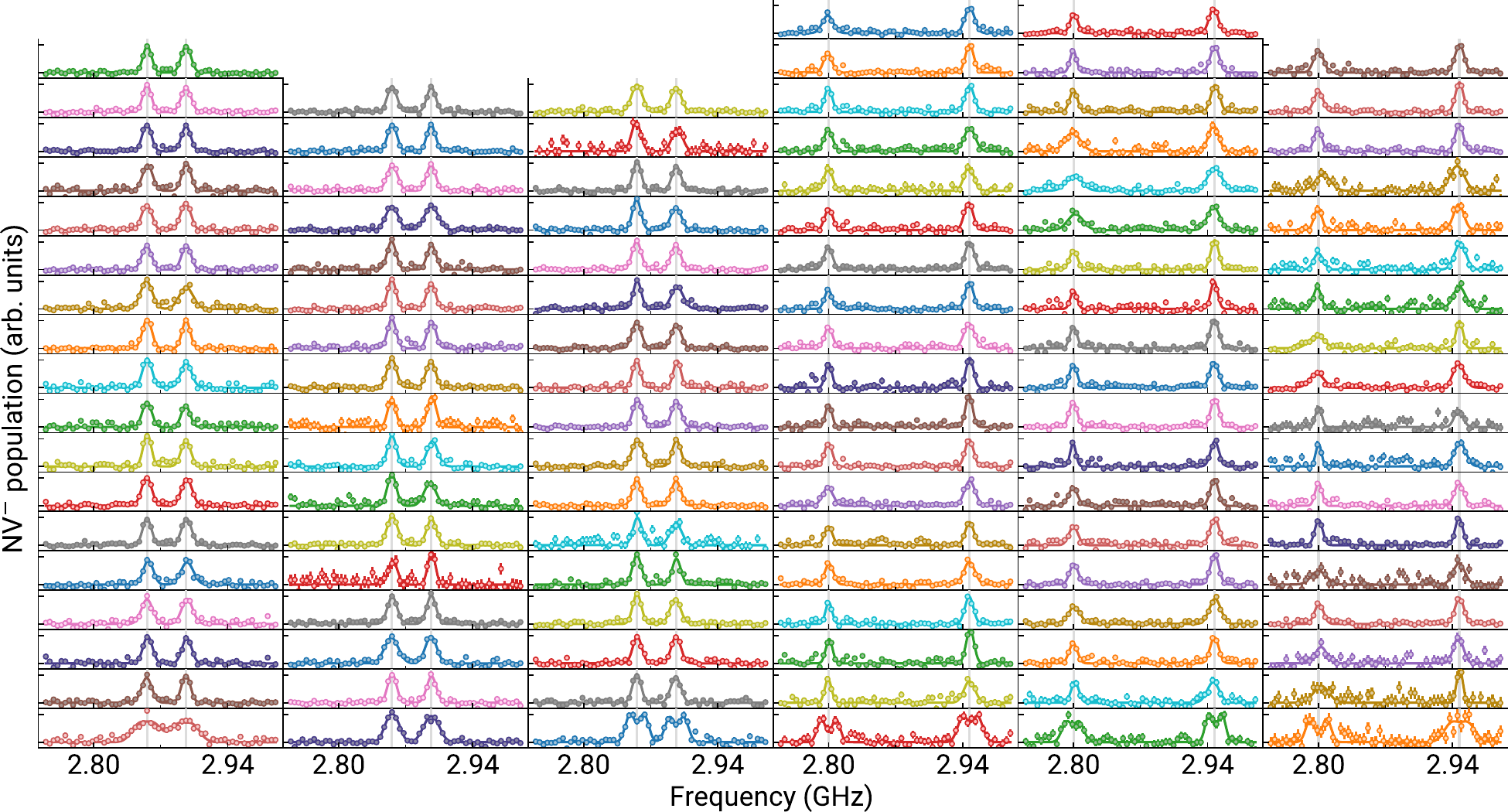}
    \caption{Parallel pulsed electron spin resonance (ESR) measurements with 108 NV centers.
    NV$^{-}$ populations are normalized such that 0 (1) corresponds to the NV$^{-}$ population measured after preparation in \(m_{s}=0\) (\(m_{s}=-1\)). Error bars (one standard error) are typically smaller than the data points.
    The left (right) three columns show traces for NV centers belonging to orientation A (B). The gray lines show the mean positions of the two resonances for each orientation. Each of the traces on the bottom row shows a significant broadening or resolved splitting, indicating the presence of a nearby \(^{13}\)C with a coupling strength on the order of 10 MHz. The total measurement time from start to finish was 7.5 hours, including normalization shots and drift tracking (Appendix~\ref{subsec:app-total_durations}).
    }
    \label{fig:parallel_measurements}
\end{figure*}

\section{Platform for parallel measurement}\label{sec:platform}

The experiments described in the main text are conducted using NV centers from a 2D layer of as-grown NV centers within a bulk diamond sample. In Appendix~\ref{sec:app-shallow}, we describe additional measurements performed using a 2D layer of implanted shallow NV centers (implantation energy of $6~$keV corresponding to a depth of approximately 10 nm \cite{pezzagna2010creation}), where we find performance comparable to the bulk diamond case. The experiments described in this work are conducted under ambient conditions. However, our approach could readily be integrated into a cryostat for low temperature experiments. Experiments are performed on 117 spots from within a 30~{\textmu}m square region in the diamond (Fig.~\ref{fig:scheme}(a-b)). 
Of the original 117 spots, 9 ultimately show weak electron spin resonance (ESR) contrast and are not included in further analysis. The remaining 108 spots are confirmed to be charge-stable single NV centers based on the structure of the histograms of integrated photon counts from each spot. 
For the green histogram shown in Fig.~\ref{fig:scheme}(c), an example NV center is prepared in NV$^{-}$ with approximately $75\%$ probability by a 520-nm charge-polarization pulse, and the charge state is subsequently read out in a single shot with the camera under 589-nm illumination. The histograms for the NV centers examined in this work are well-described by a bimodal probability distribution, indicating that all the spots studied are single NV centers (Appendix~\ref{sec:app-readout}). The two well-resolved modes correspond to the NV charge states and allow for single-shot charge-state readout by thresholding \cite{aslam2013photo}. Alternatively, the NV center can be prepared in NV$^{0}$ with an additional 638-nm ionization pulse applied after the initial 520-nm charge-polarization pulse (red histogram). The experimental apparatus is illustrated in Fig.~\ref{fig:scheme}(d). Charge-manipulation  is performed in series using 2D acousto-optic deflectors (AODs) to quickly reposition the 520- and 638-nm beams. Because charge manipulations are performed with one NV center at a time, we have the ability to selectively manipulate the charge states of certain subsets of the NV centers while leaving others undisturbed. This can be used to realize high-fidelity charge-state polarization with conditional logic (Appendix~\ref{sec:app-conditional_init}). Parallel charge-state readout is achieved using a spatial light modulator (SLM) to structure the 589-nm illumination into diffraction-limited spots targeting the individual NV centers of interest. 


\section{Parallel high-fidelity spin experiments}\label{sec:parallel}

Our platform enables independent experiments with multiple spins to be conducted in parallel using the spin-to-charge conversion (SCC) sequence shown in Fig.~\ref{fig:scheme}(e) \cite{shields2015efficient}. 
Fig.~\ref{fig:parallel_measurements} shows the resonance frequencies of 108 NV centers' \(m_{s}=0\leftrightarrow m_{s}=\pm 1\) transitions, as measured in parallel with pulsed electron spin resonance (ESR). 
Two of the possible four orientations of NV centers are represented. While all four orientations of NV centers are present in our diamond sample, we target NV centers of only two specific orientations in order to limit the number of microwave frequencies and pulses that are required to resonantly address all the NV centers in the group. For the remaining experiments described in this work, which require resonant microwave pulses, each microwave pulse consists of a pair of separate pulses with one pulse addressing each orientation and a short gap (16 ns) between the component pulses. Our platform requires that the NV center resonant frequencies are adequately homogeneous so that NV centers of the same orientation are equally addressed by the microwave pulse for that orientation. The power-broadened linewidths, which are determined by the Rabi frequency of the applied microwaves, are significantly broader than the inhomogeneities in the line centers between NV centers, indicating that all NV centers of the same orientation are driven resonantly by the same microwave tone. Specifically, for the orientation with the narrower (wider) splitting, orientation A (B), the mean full width at half maximum linewidth averaged over both resonances is 12.0 MHz (11.5 MHz) while the standard deviation of the line centers is only 0.5 MHz (0.8 MHz). 
Of the 108 NV centers shown in Fig.~\ref{fig:parallel_measurements}, 6 (bottom row) display large broadenings or splittings of their ESR lineshapes which cause their resonant or Rabi frequencies to differ significantly from other NV centers of the same orientation, which we attribute to strongly coupled ${}^{13}$C nuclei. An animated version of the ESR data set shown
in Fig.~\ref{fig:parallel_measurements} is available online (Appendix~\ref{sec:app-animations}). 

\begin{figure}[!t]
    \includegraphics[width=0.48\textwidth]{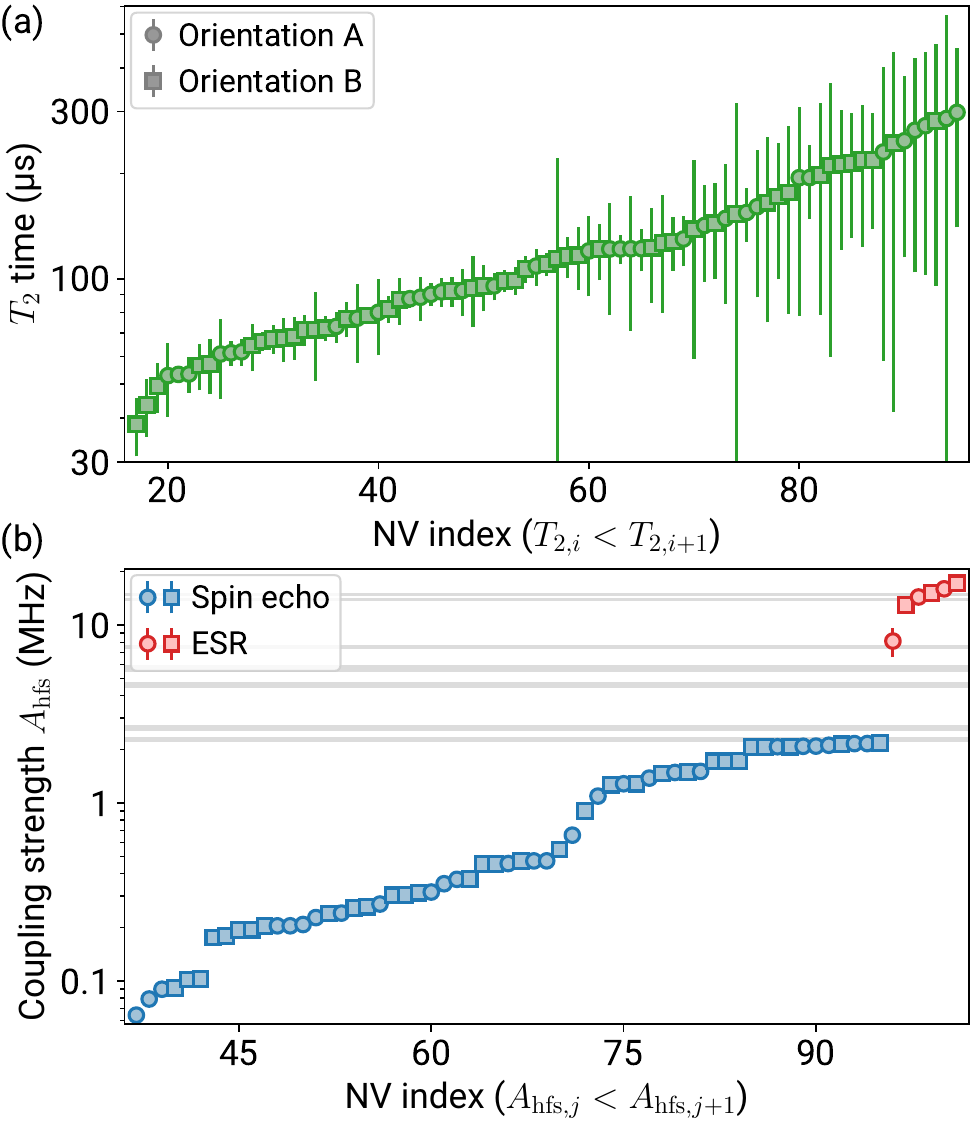}
    \caption{Parallel spin echo experiments. 
    (a) Coherence times \(T_{2}\) extracted from fits to the data from parallel spin echo experiments on 108 NV centers (see Fig.~\ref{fig:spin_echo-data} in Appendix~\ref{sec:app-spin_exp_seqs} for the raw spin echo data and fitted curves,) arranged in ascending order of measured \(T_{2}\).
    The spin echo traces for the 6 NV centers with the most strongly coupled ${}^{13}$C nuclei cannot be reliable analyzed due to microwave pulse errors from the hyperfine splittings. 
    For 6 additional NV centers the spin echo fit failed to converge. Meaningful results are therefore extracted for a total of 92 out of the 108 NV centers. The spin echo fit is consistent with no decay up to the maximum measured total evolution time of 113~{\textmu}s for 17 NV centers. These values are therefore excluded from the plot and the NV index begins at 17. 
    (b) ${}^{13}$C hyperfine coupling strengths extracted from fits to the same parallel spin echo experiments as shown in panel (a), arranged in ascending order. The spin echo data from 37 NV centers are consistent with no strongly coupled ${}^{13}$C. These values are excluded from the plot and the NV index begins at 37. For the 6 NV centers with very strongly coupled ${}^{13}$C the coupling strengths are extracted from the ESR data directly (red points). Horizontal gray regions show \textit{ab initio} hyperfine coupling strengths up to \(1\sigma\) confidence intervals for neighboring lattice sites according to the calculations in Ref.~\cite{smeltzer201113c}.
    The total measurement time from start to finish was 33 hours, including normalization shots and drift tracking (Appendix~\ref{subsec:app-total_durations}).
    }
    \label{fig:spin_echo}
\end{figure}

We next perform parallel spin echo measurements, which can be used to efficiently screen for NV centers with particular properties required by an experiment, such as long coherence times, or the presence or absence of strongly coupled ${}^{13}$C nuclei. By fitting to the resulting data, we determine the spin echo \(T_{2}\) coherence times for all of the NV centers in parallel, which we plot in ascending order of coherence in Fig.~\ref{fig:spin_echo}a. (See Appendix~\ref{sec:app-spin_exp_seqs} for further details regarding the spin echo sequence and fitting procedure.) Although the microwave pulses are applied globally, the coherent evolution of each NV center during the spin echo sequence also provides information about its unique local environment. In particular, oscillations in the spin echo traces indicate the presence of strongly coupled ${}^{13}$C nuclei \cite{childress2006coherent}, providing a proof-of-principle example of nanoscale NMR performed with multiple NV centers simultaneously. 
We attribute the large broadenings or splittings exhibited by 6 NV centers in Fig.~\ref{fig:parallel_measurements} to the presence of a strongly coupled ${}^{13}$C, yielding a coupling strength which is extracted directly by fitting to the ESR spectra. 
For an additional 6 NV centers, the spin echo fit does not converge, which we hypothesize may be due to sampling rate effects or multiple strongly coupled ${}^{13}$C nuclei in the local environments of these NV centers, which is not captured by our fitting function (Appendix~\ref{sec:app-spin_exp_seqs}). For the remaining 96 NV centers the ${}^{13}$C hyperfine coupling strengths \(A_{\text{hfs}}\) can be extracted by fitting to the spin echo traces. The fits show 37 NV centers whose spin echo traces are consistent with no coupled ${}^{13}$C. The ${}^{13}$C couplings for the remaining NV centers, as well as those extracted from ESR, are shown in Fig.~\ref{fig:spin_echo}a. Sorting the modulation frequencies in ascending order reveals terraces at particular values, which arise because the carbon nuclei exist only at discrete locations in the diamond lattice relative to the NV center. The strongest coupling strengths we observe are consistent with those identified previously as corresponding to specific neighboring lattice sites \cite{smeltzer201113c}. If necessary, more sophisticated pulse sequences could be used to detect dozens of nuclear spins coupled to each NV center in parallel \cite{cujia2022parallel, van2024mapping}.  

With the parameters used here our platform provides a speedup over fully serial measurements conducted with conventional spin readout using 532-nm illumination for experiments with 100 NV centers and interrogation times of 25~{\textmu}s or longer, where for conventional readout we assume a saturated count rate of 100 kcps, a readout duration of 200 ns, and a spin contrast of 30\% (Appendix~\ref{sec:app-speedup}).
With realistic improvements to measurement signal-to-noise ratios and readout times, we project that an order-of-magnitude speedup over serial measurements with conventional readout could be achieved for the same number of NV centers and an interrogation time of 25~{\textmu}s. (See Appendix~\ref{sec:app-speedup} for details.)

\section{Pairwise shot-to-shot spin correlations}\label{sec:correlations}

\begin{figure}[!t]
    \includegraphics[width=0.48\textwidth]{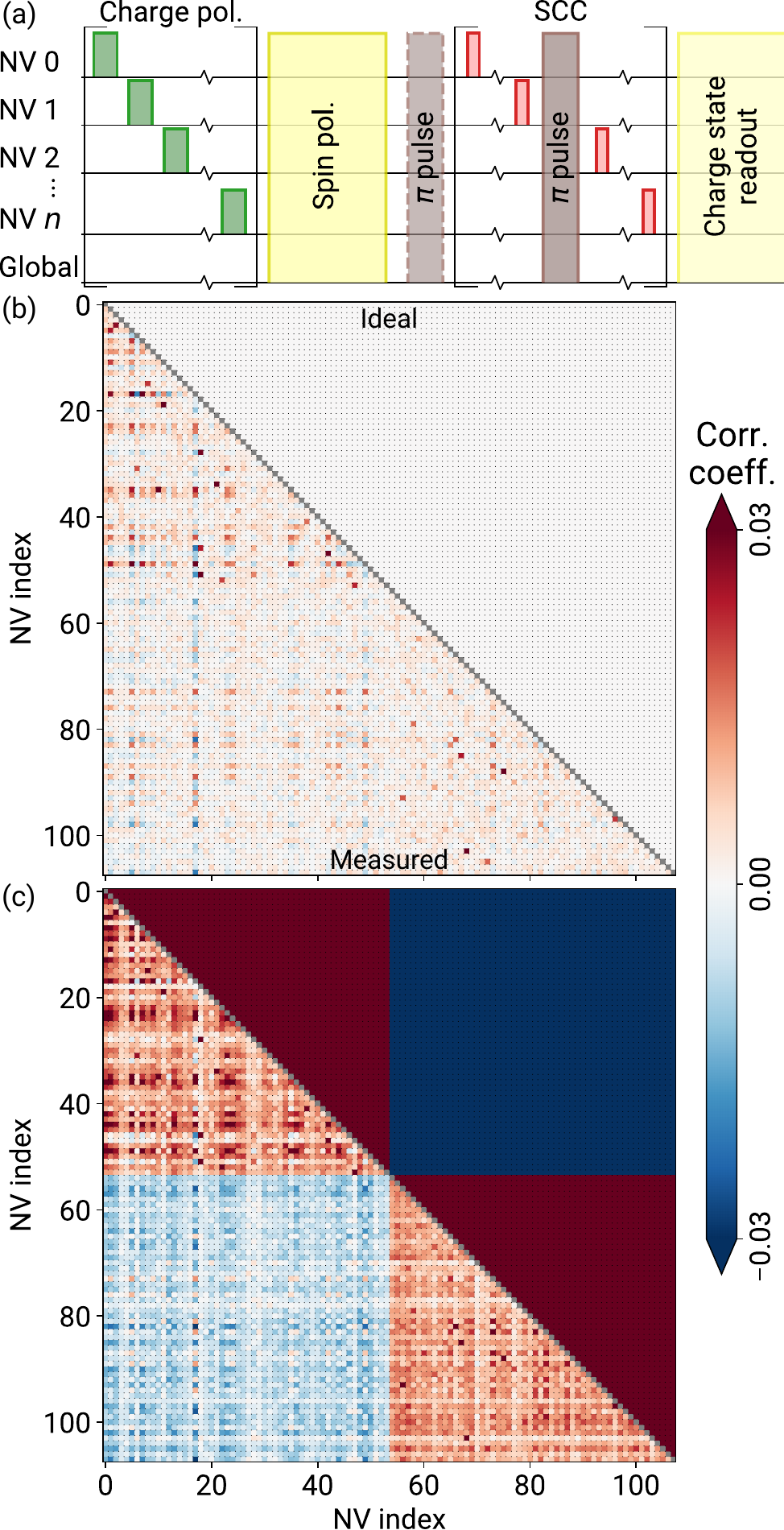}
    \caption{Pairwise correlations between spin states of 108 NV centers.
    (a) Experimental sequence. 
    Each brown block indicates a pair of microwave \(\pi\) pulses addressing the two orientations of NV centers under study. The dashed block is either applied or not applied randomly from shot to shot. 
    (b-c) Correlation matrices for reference (b) and signal (c) experiments. 
    The measured results (lower triangular sections) mirror the ideal results (upper triangular sections) in both cases. 
    Each matrix contains 5,778 unique coefficients.
    The diagonal elements, which are equal to 1 by definition, are shown in gray. 
    (b) No microwaves are applied. In the ideal case, there are no correlations. 
    (c) An additional \(\pi\) pulse inserted during SCC splits the NV centers into two groups at index 54, generating correlations (anticorrelations) between the spin states of NV centers belonging to the same group (different groups). 
    The total measurement time was 6.3 hours for each matrix, including drift tracking (Appendix~\ref{subsec:app-total_durations}).
    }
    \vspace{-25pt}  
    \label{fig:correlations}
\end{figure}

Our platform also allows for the detection of classical shot-to-shot pairwise correlations across all targeted NV center spins \cite{rovny2022nanoscale}. The number of pairwise correlation coefficients that can be accessed corresponds to the number of unique pairs of NV centers that can be formed from the complete set, which scales with the total number of NV centers \(n\) as \((n^{2}-n)/2\). For \(n=108\) NV centers, we measure 5,778 correlation coefficients. The experimental sequence used to generate and measure correlations is shown in Fig.~\ref{fig:correlations}(a). Following charge- and spin-polarization, a microwave \(\pi\) pulse is randomly either applied or not applied from shot to shot, setting up positive correlations between the spin states of all 108 NV centers. Negative correlations can then be introduced by inserting an additional \(\pi\) pulse within the serialized SCC step. This additional \(\pi\) pulse causes the spins of NV centers which have not yet undergone SCC to flip, becoming anticorrelated with the spins of those that have already undergone SCC. Here we arrange the SCC sequence and the additional \(\pi\) pulse to divide the NV centers spatially into two groups of equal number (Appendix~\ref{subsec:dividing_line}). The order in which the two groups undergo SCC is swapped every other shot of the experiment to even out contrast loss associated with spin relaxation. 

We first conduct the correlation measurement sequence with no microwave pulses applied (Fig.~\ref{fig:correlations}(b)). Ideally, there would be no shot-to-shot correlations in this case. In practice, fluctuations in laser intensities, sample drift, and noise sources affecting the EMCCD camera all introduce background correlations in the measured number of photons between NV centers. 
The mean background correlation level we measure is \(0.0015 \pm 0.0040\) (one standard deviation).
Upon introduction of the randomly applied \(\pi\) pulse (Fig.~\ref{fig:correlations}(c)), we measure correlations with a mean absolute value of \(0.0091 \pm 0.0060\), six times larger than the background level.
We note that in most covariance sensing applications quantum projection noise will decrease the contrast of the measured correlations by a factor of 2 relative to the induced correlations measured here. 

The results presented in Fig.~\ref{fig:correlations} demonstrate that our platform meets three important requirements for practical covariance magnetometry with NV centers \cite{rovny2022nanoscale}. First, high-fidelity single-NV measurements are required, as the amount of time needed to resolve a correlated signal scales with the fourth power of the readout noise of a single NV center \cite{rovny2022nanoscale}, rather than with the square of the readout noise as in the case of non-correlated measurements \cite{degen2017quantum}. Notably, the readout noise for parallel measurements of multiple NV centers with an EMCCD is similar to previously demonstrated values using a confocal microscope to interrogate a single NV center with SCC (Appendix~\ref{sec:app-scalability}). Second, the number of correlation coefficients probed in a single experiment scales combinatorially with the number of NV centers measured simultaneously. Practical covariance magnetometry therefore benefits from the parallel operation of many sensors in order to probe a range of correlation lengths, rather than measuring one two-point correlation function at a time. 
Finally, our ability to generate arbitrary pairwise anticorrelations by flipping a subset of NV spins prior to readout can be used to isolate spin correlations by canceling out common-mode contributions from background technical correlations \cite{rovny2022nanoscale}. 

\section{Outlook}\label{sec:outlook}

We have demonstrated parallel control and measurement of the electronic spin states of 108 individual NV centers simultaneously. 
The number of NV centers addressed simultaneously was primarily limited by background fluorescence during charge-state readout. 
Background fluorescence impacts the duration of the exposure required to achieve single-shot charge state readout, and thereby affects total measurement time. As we attribute most of the background fluorescence we observe to out-of-focus NV centers at other depths within the diamond, the background level will scale roughly linearly with the number of NV centers under examination. This sets up a tradeoff between the number of NV centers under interrogation and the measurement time. Here we opted to conduct measurements with \(\mathcal{O}(100)\) NV centers and a 50-ms readout duration. There are several approaches by which the background level could be reduced in future work. 
Illumination with a light sheet would eliminate background from out-of-focus sources \cite{stelzer2021light}. A delta-doped high-purity diamond sample could offer the same benefits as a light sheet without requiring modifications to the experimental apparatus \cite{ohno2012engineering}. 

With reduced background fluorescence, we expect the scalability of our approach to become limited by the spin lifetimes of the NV centers. At room temperature, the spin state relaxes on millisecond time scales for NV centers deep in bulk diamond \cite{cambria2021state}, and shorter time scales for shallow NV centers \cite{myers2017double, sangtawesin2019origins} and NV centers in nanodiamonds \cite{gardill2020fast}. Given the access times of the AODs used in this work, around 500 NV centers could be addressed in series within the $1/e$ time constant for spin contrast at room temperature (Appendix~\ref{sec:app-scalability}). Faster AOD operation could allow for parallel experiments with over 1,000 NV centers at room temperature. At cryogenic temperatures, where the spin lifetime for NV centers deep in high-purity bulk diamond can exceed 10 seconds \cite{cambria2023temperature}, we anticipate that scalability will be limited by the angular bandwidth of the AODs. In this regime we estimate that parallel measurements with over 9,000 NV centers should be achievable, with the ultimate limit determined by optical crosstalk effects (Appendix~\ref{sec:app-crosstalk}) and the field of view of the optics (Appendix~\ref{sec:app-scalability}). For shallow NV centers, noise sources associated with the surface may persist at low temperatures and reduce these numbers. Measurements performed at cryogenic temperatures could also take advantage of single-shot spin-state readout techniques to further enhance the signal-to-noise ratio and enable conditional spin-state preparation \cite{robledo2011high, irber2021robust, zhang2021high}.

In the near term, we believe that our approach to scalable spatial multiplexing will broaden the applications of nanoscale quantum sensing, enabling sensitive, high spatial resolution measurements of systems that exhibit weak signals or require the collection of large data sets for statistics \cite{du2024single}, and the characterization of materials that feature complex spatial correlations over micron length scales \cite{kolkowitz2015probing, jenkins2019single}. 
Parallelized operation will also accelerate quantum information processing and quantum simulation with nuclear spin registers \cite{randall2021many, abobeih2022fault}. Lastly, we note that the techniques developed here can be readily adapted for use with other solid state spin systems, such as defects in silicon carbide \cite{luo2023fabrication} and hexagonal boron nitride \cite{vaidya2023quantum}.

\vspace{0.3cm}
\noindent \textbf{\textit{Authors' note:}} We are aware of a manuscript that describes complementary work performed in parallel to our own in which multiplexing of measurements with individual NV centers in diamond is demonstrated \cite{ChengParallel}. Both works deploy low readout-noise cameras and SLMs to measure many NV centers simultaneously; in Ref.~\cite{ChengParallel}, Cheng \textit{et al.} perform all operations in parallel and use the same illumination source for both SCC and charge-state readout. 

\section*{Acknowledgments}

We thank Carter Fox for contributions to the early stages of the experiment, and Kevin Villegas for suggestions regarding the experimental sequences. We acknowledge helpful conversations with Jared Rovny, Jeff Thompson, Kai-Hung Cheng, Zeeshawn Kazi, and Nathalie de Leon. Design and construction of the experimental apparatus, as well as work on parallel ESR and spin echo measurements, was supported by the U.S.~Department of Energy Office of Science National Quantum Information Science Research Centers as part of the Q-NEXT center. Work on high fidelity charge state readout and conditional charge state initialization was supported by the U.S.~Department of Energy, Office of Science, Basic Energy Sciences under Award No.~DE-SC0020313. Work on parallel shot-to-shot correlation measurements was supported by the NSF under Award No.~2326767.

\setcounter{section}{0}
\renewcommand{\theHsection}{\Alph{section}}
\renewcommand{\thesection}{\Alph{section}}
\titleformat{\section}[block]{\normalfont\bfseries}{APPENDIX \thesection.}{1em}{\centering}
\setcounter{subsection}{0}
\renewcommand{\theHsubsection}{\arabic{subsection}}
\renewcommand{\thesubsection}{\arabic{subsection}}

\section{Performance with shallow NV centers}\label{sec:app-shallow}

\begin{figure*}[!t]
    \includegraphics[width=1\textwidth]{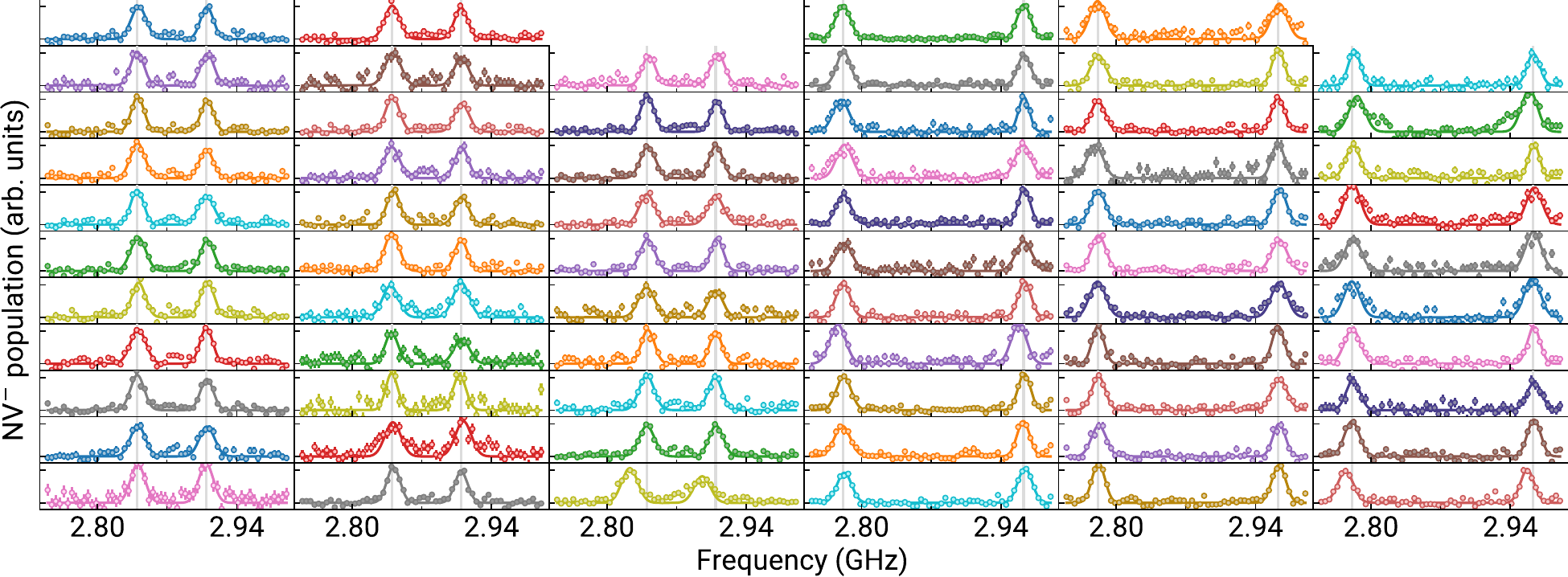}
    \caption{Parallel ESR with 64 shallow NV centers. 
    As in Fig.~\ref{fig:parallel_measurements}, NV$^{-}$ populations are normalized such that 0 (1) corresponds to the NV$^{-}$ population measured after preparation in \(m_{s}=0\) (\(m_{s}=-1\)) (Appendix~\ref{subsec:app-normalization}). Error bars are one standard error.
    The left (right) three columns show traces for NV centers belonging to the same orientation. The gray lines show the mean positions of the two resonances for each orientation, excluding the final 2 NV centers from each orientation group, which show significant shifts in their resonances. 
    The total measurement time from start to finish was 5.6 hours, including normalization shots and drift tracking (Appendix~\ref{subsec:app-total_durations}).
    }
    \label{fig:shallow_esr}
\end{figure*}

\begin{figure}[!t]
    \includegraphics[width=0.48\textwidth]{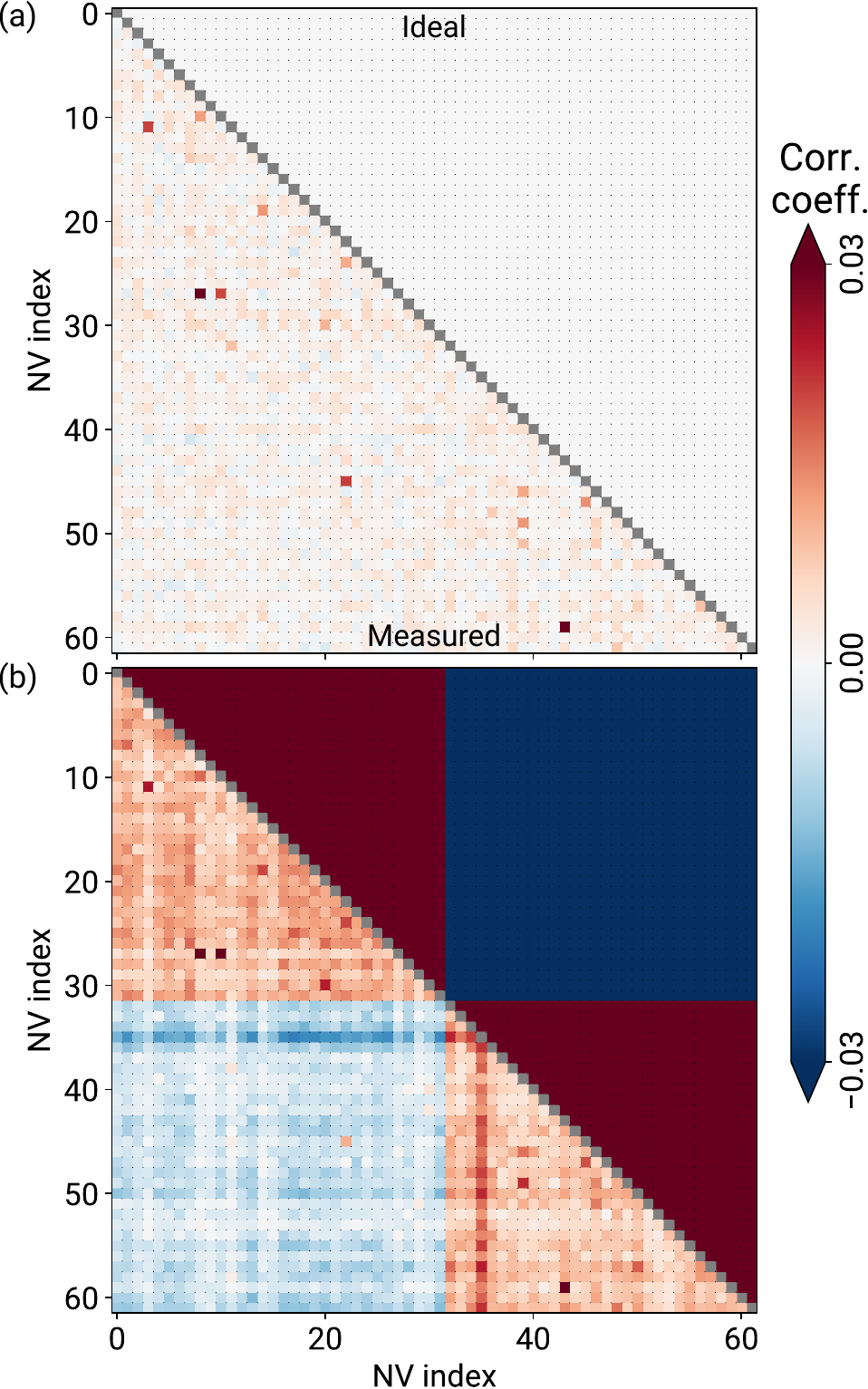}
    \caption{Pairwise correlations between spin states of 62 shallow NV centers. The 2 NV centers with large resonance shifts are not included in the plot. 
    The measurement sequence is the same as in Fig.~\ref{fig:correlations}. The additional \(\pi\) pulse divides the NV centers into two groups at index 32. 
    The total measurement time was 7.8 hours for each matrix, including drift tracking (Appendix~\ref{subsec:app-total_durations}).
    }
    \label{fig:shallow_correlations}
\end{figure}

In addition to the measurements described in the main text, which are performed with NV centers approximately 15~{\textmu}m below the surface of Sample I, a bulk diamond sample, we also perform ESR (Fig.~\ref{fig:shallow_esr}) and correlation (Fig.~\ref{fig:shallow_correlations}) measurements with a second sample, Sample II, which was implanted with NV centers at an expected average depth of around 10 nm based on the implantation energy used (6 keV) \cite{pezzagna2010creation}. (See Appendix~\ref{subsec:app-samples} for sample details.) Experiments are conducted on 69 spots, of which 64 show strong ESR contrast. In the ESR data, there is a small amount of additional inhomogeneity in the microwave transition frequencies. Specifically, for the orientation with the narrower (wider) splitting, the mean full width at half maximum linewidth is 16.7 MHz (17.0 MHz) while the standard deviation of the line centers is 0.7 MHz (1.4 MHz). These values exclude the final two NV centers from each group, which display especially large shifts, consistent with the effects of strain or electric fields from charge traps on the diamond surface. The power-broadened linewidths are still an order of magnitude larger than the inhomogeneity, indicating that NV centers of the same orientation can be equivalently driven by the same microwave pulse. For the correlation experiment the NV centers are divided into two groups spatially along the dividing line shown in Fig.~\ref{fig:dividing_line}(b). There are 32 NV centers in one group and 30 in the other. In the correlation data, the mean background level (Fig.~\ref{fig:shallow_correlations}(a)) is \(0.0013\pm0.0025\) (one standard deviation), while the mean absolute value of the signal data (Fig.~\ref{fig:shallow_correlations}(a)) is \(0.0075\pm0.0041\), nearly 6 times higher than the background level. The magnitude of the measured spin correlations is similar to that measured with deep NV centers in Sample I.

These results demonstrate that our platform can be used with shallow NV centers with minimal performance degradation. To support this claim, we additionally note that prior work introducing NV center correlation measurements was conducted with two shallow NV centers each around 10 nm below the surface \cite{rovny2022nanoscale}. The results reported in all other sections of this work pertain to the measurements conducted in Sample I unless otherwise noted.

\section{Wide-field version of apparatus}\label{sec:app-widefield}

An earlier version of the experimental apparatus used wide-field 589-nm illumination rather than patterned 589-nm illumination with an SLM. The wide-field source was achieved by focusing a single 589-nm beam at the back aperture of the objective, creating a spot of roughly 20~{\textmu}m in diameter at the objective focal plane. In comparison to the SLM-based version of the apparatus, the wide-field version allowed for the same measurements to be conducted, but the scalability was limited to around 10 NV centers for a 50-ms readout due to background fluorescence. The scalability of the wide-field approach is more significantly limited by background because much of the intensity of the wide-field beam is not incident on any NV center and therefore only contributes background without providing any signal. The experiments discussed in 
Appendices~\ref{sec:app-conditional_init} and \ref{subsec:app-optical_crosstalk} 
were conducted using the wide-field version of the apparatus. In addition, these experiments used linearly polarized light rather than circularly polarized light. The linear polarization was rotated to optimally excite two specific NV center orientations of the four supported by the diamond. 
The results described in the rest of this work pertain to measurements conducted using the SLM-based readout approach unless otherwise noted.

\section{Experimental details}\label{sec:app-exp_details}

The EMCCD camera used in this work is a N{\"u}v{\"u} HN{\"u} 512 Gamma with a resolution of \(512 \times 512\) pixels. A region of interest of \(512 \times 200\) pixels is used to reduce the camera readout duration. A Quantum Machines OPX is used for experimental sequencing, including conditional logic and AOD drive tone generation. As NV centers are targeted one at a time by changing the AOD drive tones, only two analog output channels from the OPX are required per AOD. The AODs used in this work are Brimrose 2DS series. While in principle a single 2D AOD could be used to control both the 520- and 638-nm beams if the beams are overlapped before the AOD, for this work we opted to decouple the 520- and 638-nm beams and use a dedicated 2D AOD for each beam. The SLM is a Thorlabs Exulus-HD2. The excitation beams are circularly polarized using an achromatic quarter waveplate (Newport 10RP54-1B) placed before the collection dichroic. 
The collection dichroic used to split the excitation wavelengths from NV center photoluminescence is a longpass dichroic with an edge at 640 nm. The 4\(f\) system relaying the excitation light from the AODs and SLM to the back aperture of the objective consists of two achromatic doublets. The first lens has a focal length of 18 cm and the second lens has a focal length of 30 cm. 

The microscope objective is an Olympus MPLAPON100XO2 oil objective with a numerical aperture of 1.45 and a back aperture diameter of 5.2 mm. The 520-nm, 589-nm, and 638-nm beams are sized to fill the back aperture after magnification through the telescope in the 4\(f\) relay. A permanent magnet is used to lift the degeneracy between the \(m_{s}=\pm1\) spin states and allow for microwave frequency resolution of NV centers of different orientations. The magnitude of the applied magnetic field is approximately 37 G. We study NV centers belonging to two orientations out of the four supported by the diamond. The ESR traces shown in Fig.~\ref{fig:parallel_measurements} are grouped by orientation, with orientation A (B) traces comprising the left (right) three columns of the plot. The NV axis of orientation A (B) makes an angle of 79\degree (45\degree) with the applied magnetic field. Microwave pulses are delivered by the antenna described in Ref.~\cite{sasaki2016broadband} with a Rabi frequency of roughly 10 MHz. The diamond sample is mounted directly on the antenna which is itself mounted on a 3-axis piezo stage (PI P616 NanoCube). The piezo stage is used to correct for physical drift of the sample with respect to the objective roughly once per minute.

The 520-nm laser is a direct diode Integrated Optics Matchbox. The 638-nm laser is a direct diode Cobolt 06-01. The 520- and 638-nm lasers are digitally modulated directly from the OPX. The 589-nm laser is a DPSS model from Opto Engine. The 589-nm laser is modulated externally with an acousto-optic modulator (AOM) from Isomet. The AOM features analog modulation which is used to adjust the power of the 589-nm light for the spin polarization and charge-state readout pulses.

We drive global spin rotations equally for both NV orientations using a pair of back-to-back microwave pulses at different frequencies. The pulse pairs are generated by two microwave signal generators (SRS SG394), with a dedicated signal generator for each NV orientation. Each signal generator is gated using a discrete microwave switch, with the switches wired in series such that the output line of the second switch carries the pulses from both second generators. In this scheme, a \(\pi\) pulse consists of the sequence \(\pi_{A}-\pi_{B}\) where \(\pi_{A}\) (\(\pi_{B}\)) denotes a microwave \(\pi\) pulse resonant with orientation A (B) NV centers. A short gap of 16 ns separates the two component pulses. \(\pi/2\) pulses are delivered analogously. 

\subsection{Diamond samples}\label{subsec:app-samples}

The bulk diamond sample (Sample I) used for the measurements shown in the main text of this work is from Element Six and was grown by chemical-vapor-deposition (CVD). The surface orientation is 100. The NV centers in Sample I were formed naturally during growth. The sample displays an inhomogeneous concentration of NV centers with different areas exhibiting concentrations between \(10^{-5}\) and \(10^{-3}\) ppb, as estimated by direct counting. The higher-concentration regions are localized in irregularly-spaced layers that are nearly two-dimensional relative to the microscope depth of field. The NV centers studied in Sample I are from the second such plane from the surface, about 15~{\textmu}m below the surface. The NV centers in this plane allow for the highest fidelity single-shot charge-state readout. Aberrations associated with the high index of refraction of diamond diminishes readout fidelity for NV centers in planes further from the surface. A slight (roughly 10\%) increase in background fluorescence diminished readout fidelity for the NV centers in the first plane from the surface, about 5~{\textmu}m down. 

The shallow-implanted diamond sample (Sample II) was purchased from Applied Diamond and was also grown by CVD. The surface orientation is 100. Sample II was implanted with \(^{15}\)N at a flux of \(10^{9}\) cm\(^{2}\) and an energy of 6 keV, corresponding to a depth of roughly 10 nm \cite{pezzagna2010creation}. Following implantation, NV centers were formed by annealing. 

\subsection{NV center identification}\label{subsec:app-identification}

Potential NV centers are identified by raster scanning the 520-nm beam over small (roughly \(10\times10\)~{\textmu}m) patches of the diamond while exposing the camera. Potential NV centers are then identified as bright spots within each image. This approach is taken, as opposed to raster scanning the 520-nm beam over the full region of interest during a single exposure, in order to minimize background and increase the signal-to-noise ratio of the NV centers in the images. The composite wide-field image shown in Fig.~\ref{fig:scheme}(a) shows the maximum value of each pixel recorded across the 9 component images that were recorded to cover the full region of interest. Taking the maximum obscures some NV centers which are otherwise visible in the component images, and which are visible in the single-shot image under patterned illumination shown in Fig.~\ref{fig:scheme}(b).

\subsection{Optical pulse parameters}\label{subsec:app-pulse_parameters}
520-nm charge polarization pulses are 1~{\textmu}s at 5 mW. 638-nm ionization pulses (used for charge state histograms shown in Figs.~\ref{fig:scheme} and \ref{fig:histograms-supp}) are 1~{\textmu}s at 5 mW. 638-nm SCC pulses are optimized in both duration and power for each NV center individually, but are 60--300 ns at 9--10 mW. 589-nm spin polarization pulses are 10~{\textmu}s at 0.3 mW. 589-nm charge-state readout pulses are 50 ms. The readout power for each NV center is tuned individually, but is 1-4~{\textmu}W. The total power of the 589-nm light for the experiments with 108 NV centers is 234~{\textmu}W. The 100-\textmu s spin-polarization pulse is applied using the same SLM pattern as for readout but at a higher total power of approximately 900~{\textmu}W.

\subsection{Spin experiment normalization}\label{subsec:app-normalization}

The measured NV\(^{-}\) populations for the ESR (Figs.~\ref{fig:parallel_measurements} and \ref{fig:shallow_esr}) and spin echo (Figs.~\ref{fig:spin_echo} and \ref{fig:spin_echo-data}) experiments are normalized against reference levels for each NV center where 0 (1) is defined as the NV\(^{-}\) population following initialization in the \(m_{s}=0\) (\(m_{s}=-1\)) spin state. The normalization values are measured alongside the signal values by interleaving signal shots (where the parameter of interest is varied, e.g., the microwave frequency in ESR) and normalization shots. The spin state measured in the normalization shot is alternated between \(m_{s}=0\) and \(m_{s}=-1\). The measurement consists of the same sequence shown in Fig.~\ref{fig:scheme}e used for the signal shots. 
For the 4 ESR traces which show resolved splittings in Fig.~\ref{fig:parallel_measurements} (last 4 traces on bottom row) and the 2 traces which show large shifts in Fig.~\ref{fig:shallow_esr} (third and sixth traces on bottom row) this normalization scheme works poorly as the microwave pulse used for normalization is detuned from the resonance peak. For these traces, 1 is defined as the highest measured value in the trace.

\begin{figure}[!b]
    \includegraphics[width=0.48\textwidth]{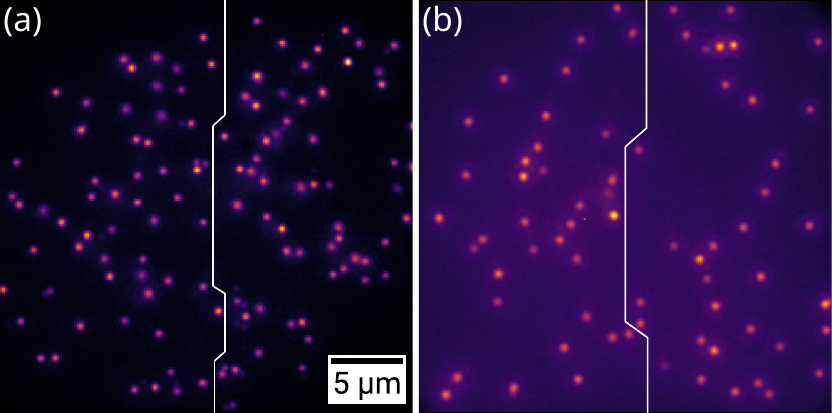}
    \caption{The white line down the center of each image demarcates the two groups of NV centers used for the correlation experiments shown in Figs.~\ref{fig:correlations} (Sample I, shown in panel (a)) and \ref{fig:shallow_correlations} (Sample II, shown in panel (b)). The scalebar is the same for both images.
    }
    \label{fig:dividing_line}
\end{figure}

\subsection{Number of shots and experimental run times} \label{subsec:app-total_durations}
Here we describe the averaging statistics for the data sets presented in this work. 
The duration for a single shot of each experiment 
is dominated by the 50-ms exposure for readout. There is a 12 ms dead time between shots as the image is read off the EMCCD pixel array, and an additional overhead of approximately 10\% associated with drift tracking. In total, a single shot takes approximately 70 ms to complete. Each point in the ESR data recorded with Sample I (Fig.~\ref{fig:parallel_measurements}) is the average of 3200 shots, and there are 60 data points in all. The total measurement time is 7.5 hours including normalization shots. Each point in the spin echo data shown in Fig.~\ref{fig:spin_echo} is the average of 9600 shots, and there are 88 data points in total. The total measurement time is 33 hours including normalization shots. Each correlation matrix recorded with Sample I (Fig.~\ref{fig:correlations}) is the average of 320,000 shots and took roughly 6.3 hours to record. Each point in the ESR data recorded with Sample II (Fig.~\ref{fig:shallow_esr}) is the average of 2400 shots, and there are 60 data points in all. The total measurement time is 5.6 hours including normalization shots. Each correlation matrix recorded with sample II (Fig.~\ref{fig:shallow_correlations}) is the average of 400,000 shots and took roughly 7.8 hours to record.

\subsection{Spatial sorting of NV centers into groups for correlation measurements}\label{subsec:dividing_line}

Fig.~\ref{fig:dividing_line} depicts the spatial positions of the two groups of NV centers used for the correlation experiments shown in Figs.~\ref{fig:correlations} and ~\ref{fig:shallow_correlations}. In the experiments, NV centers on the same (opposite) sides of the white dividing line are correlated (anticorrelated). 

\section{Charge-state readout}\label{sec:app-readout}

\subsection{Image processing and count integration} As our EMCCD is not photon number-resolving, the pixel values returned by the camera upon readout are expressed as a noisy linear function of photon number in terms of analog-to-digital converter units (ADUs). In ADUs, the image features a non-zero baseline value which corresponds to zero photons. We found that this baseline value drifts over minute timescales, which can introduce spurious correlations by adding a common-mode contribution to the apparent brightness of the NV centers in an image from shot to shot. To account for this we measure the baseline value for each image by masking off a portion of the EMCCD sensor with an iris so that the masked portion receives no light. The measured baseline is then subtracted off the image and the pixel values are converted from ADUs to approximate photon number by multiplying by a fixed factor determined from the EMCCD gain specifications. To obtain the integrated counts for an NV center we sum the approximate photon numbers for pixels whose centers are within a 2.5 pixel radius of the center of the NV center. 

The charge state of an NV center is inferred by thresholding on the integrated counts. For each experiment performed in this work besides the conditional initialization experiment (Appendix~\ref{sec:app-conditional_init}), the threshold is determined in post-processing by Otsu's method using the histogram of integrated counts obtained from the experiment itself. By performing the threshlding in post-processing we account for effects such as slow laser power drift and changes in the background count level associated with the particular microwave sequences for different experiments. For the conditional initialization experiment shown in Fig.~\ref{fig:conditional_init}, the threshold values were determined in advance. 

\begin{figure*}[!t]
    \includegraphics[width=0.98\textwidth]{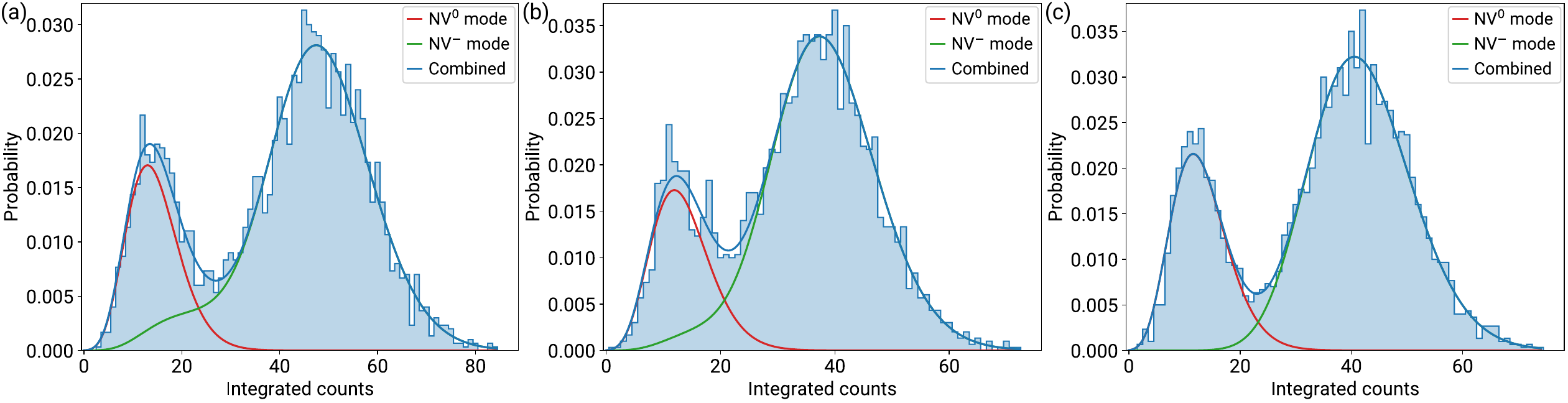}
    \caption{Charge-state histogram fitting.
    Histogram of integrated counts for three NV centers following preparation in NV\(^{-}\) by a 520-nm charge-polarization pulse. 
    The histogram is fit to the distribution given by Eq.~\eqref{weighted_prob}. The contribution to the total fit from the NV\(^{-}\) and NV\(^{0}\) modes are shown in red and green respectively, and the combined fit is shown in blue. 
    From the fits, we can extract the NV\(^{-}\) initialization fidelities and ionization rates. 
    (a) Initialization fidelity: \(P\left(X=\text{NV}^{-}\right) = 0.78\); 
    ionization rate: 3.5 s\(^{-1}\).
    (b) Initialization fidelity: \(P\left(X=\text{NV}^{-}\right) = 0.78\); 
    ionization rate: 1.9 s\(^{-1}\).
    (c) Initialization fidelity: \(P\left(X=\text{NV}^{-}\right) = 0.73\); 
    ionization rate: 0.0 s\(^{-1}\).
    }
    \label{fig:histograms-supp}
\end{figure*}

\begin{widetext}
\subsection{Charge-state histogram fitting} Several experimental figures of merit can be determined by fitting to the charge-state histograms. Here we derive a model that can be used to fit the histograms. 
We assume that the region associated with an NV center on the camera accumulates photoelectrons at a rate \(\lambda_{0}\) or \(\lambda_{-}\) according to whether the NV center is in the NV\(^{0}\) or NV\(^{-}\) charge state. 
In addition, we assume that the readout light drives NV\(^{-}\)\(\rightarrow\)NV\(^{0}\) ionization at a rate \(\gamma\), and that there is negligible probability of the NV\(^{0}\)\(\rightarrow\)NV\(^{-}\) recombination process, as 589 nm is below the NV\(^{0}\) zero-phonon line. In this model, for some fixed readout duration \(t_{r}\) the photoelectron statistics are distributed according to a Poisson distribution for NV\(^{0}\) and a Markov-modulated Poisson distribution for NV\(^{-}\). The probability distributions are
\begin{align}
    P\left(Y=y|X=\text{NV}^{0}\right) &= f_{\text{Pois}}(y;\lambda_{0}t_{r})\label{nv0_prob_inital}\\
    P\left(Y=y|X=\text{NV}^{-}\right) &= \int_{t_{r}}^{\infty}dt \,f_{\text{Exp}}(t;\gamma)f_{\text{Pois}}(y;\lambda_{-}t_{r})\nonumber\\
    &\quad + \sum_{y^{\prime}=0}^{y}\int_{0}^{t_{r}}dt \,f_{\text{Exp}}(t;\gamma)f_{\text{Pois}}(y^{\prime};\lambda_{-}t)f_{\text{Pois}}(y-y^{\prime};\lambda_{0}(t_{r}-t))\label{nvn_prob_inital}
\end{align}
where the random variable \(X\) describes the NV charge state at the start of readout and the random variable \(Y\) describes the total number of photoelectrons accumulated over the course of the readout. The function \(f_{\text{Pois}}(k;\lambda)=\lambda^{k}\exp(-\lambda)/k!\) is the probability mass function of a Poisson distribution with parameter \(\lambda\) evaluated at value \(k\) and the function \(f_{\text{Exp}}(x;\lambda)=\lambda \exp(-x\lambda)\) is the probability density function of an exponential distribution with parameter \(\lambda\) evaluated at \(x\). The first term in Eq.~\eqref{nvn_prob_inital} describes events where the NV center is not ionized during readout and all \(y\) photoelectrons come from NV\(^{-}\). The second term describes events where the NV center is ionized at some point prior to the end of readout such that some number of photoelectrons \(y^{\prime}\) is contributed by NV\(^{-}\) and the rest by NV\(^{0}\). The integral in the first term and the sum over \(y^{\prime}\) in the second term in Eq.~\eqref{nvn_prob_inital} can be evaluated, yielding
\begin{align}
    P\left(Y=y|X=\text{NV}^{-}\right) &= \left(1-e^{-\gamma t_{r}}\right)f_{\text{Pois}}(y;\lambda_{-}t_{r})\nonumber\\
    &\quad + \int_{0}^{t_{r}}dt \,f_{\text{Exp}}(t;\gamma)f_{\text{Pois}}(y;\lambda_{-}t+\lambda_{0}(t_{r}-t)).\label{nvn_prob_eval}
\end{align}

At high electron-multiplying gain, the dominant noise source affecting the readout of the photoelectrons is the stochastic nature of the electron multiplication process. Specifically, the number of electron avalanches that occurs during readout is itself Poisson distributed with mean equal to the number of photoelectrons. The number of avalanches that occurs, which corresponds to the integrated counts value described in the previous section is therefore distributed as the compound Poisson distribution
\begin{align}
    P\left(Z=z|X=x\right) &= \sum_{y=0}^{\infty}f_{\text{Pois}}(z;y)P(Y=y|X=x)\label{compound_poisson}
\end{align}
where the random variable \(Z\) describes the number of electron avalanches. The compound Poisson distribution can be well-approximated by a negative binomial distribution with probability mass function \(f_{\text{NB}}(k;r,p)=C(k+r-1, k)(1-p)^{k}p^{r}\) evaluated at \(k\) with parameters \(r\) and \(p\). The approximation is realized when \(p\) is fixed to \(p=1/2\). We abbreviate this distribution as \(f^{\prime}_{\text{NB}}(k;r)=f_{\text{NB}}(k;r,p=1/2)\) In this case the mean of the distribution is equal to the parameter \(r\) is the same as the mean of the compound Poisson distribution being approximated. Making the approximation,
\begin{align}
    P\left(Z=z|X=\text{NV}^{0}\right) &\approx f^{\prime}_{\text{NB}}(z;\lambda_{0}t_{r})\label{nv0_prob_neg_binom}\\
    P\left(Z=z|X=\text{NV}^{-}\right) &\approx \left(1-e^{-\gamma t_{r}}\right)f^{\prime}_{\text{NB}}(z;\lambda_{-}t_{r})\nonumber\\
    &\quad + \int_{0}^{t_{r}}dt \,f_{\text{Exp}}(t;\gamma)f^{\prime}_{\text{NB}}(z;\lambda_{-}t+\lambda_{0}(t_{r}-t)).\label{nvn_prob_neg_binom}
\end{align}
The integral in the second term in Eq.~\eqref{nvn_prob_neg_binom} can be calculated numerically. Each histogram is then ultimately fit to the weighted sum
\begin{align}
    P(Z=z) &= \left(1-P\left(X=\text{NV}^{-}\right)\right)P\left(Z=z|X=\text{NV}^{0}\right)\nonumber\\
    &\quad + P\left(X=\text{NV}^{-}\right)P\left(Z=z|X=\text{NV}^{-}\right)\label{weighted_prob}
\end{align}
where the fit parameters are: the probability that the NV center began readout in NV\(^{-}\), \(P(X=\text{NV}^{-})\); the mean integrated counts from NV\(^{0}\), \(\lambda_{0}t_{r}\); the mean integrated counts from NV\(^{-}\), \(\lambda_{-}t_{r}\); and the NV\(^{-}\)\(\rightarrow\)NV\(^{0}\) ionization rate, \(\gamma\).

Example histograms and fits for 3 of the NV centers from the group of 108 described in the main text are shown in Fig.~\ref{fig:histograms-supp}. The histograms shown are recorded following charge-polarization into NV\(^{-}\). Fitting to histograms from all 108 NV centers yields an average NV\(^{-}\) initialization fidelity of 0.78 and an average NV\(^{-}\)\(\rightarrow\)NV\(^{0}\) ionization rate of 2.1 s\(^{-1}\). For the 50-ms readout duration this corresponds to ionization probability 10\%. The model described by Eq.~\eqref{weighted_prob} provides an excellent fit to the charge state histograms for all the NV centers examined in this work. This indicates that the NV centers are single NV centers, as groups of NV centers within a single diffraction-limited spot would in general display a histogram with more than two modes, corresponding to different numbers of NV centers in NV\(^{-}\) and NV\(^{0}\). 
\end{widetext}

\section{Pulse sequences and fit details for ESR and spin echo experiments}\label{sec:app-spin_exp_seqs}

\subsection{ESR}

The microwave sequence for the ESR data shown in Fig.~\ref{fig:parallel_measurements} consists of a single 64-ns microwave pulse of varying frequency. The fit function is a sum of two Gaussian profiles with different center frequencies, linewidths, and amplitudes. For the 6 NV centers which evidence a strongly coupled \(^{13}\)C (bottom row of Fig.~\ref{fig:parallel_measurements}), we include an additional splitting parameter, which is the same for both resonances. The lineshape for each resonance in this case is a sum of two Gaussians which are offset in opposite directions from the resonance center frequency by half the splitting.

\subsection{Spin echo}

The microwave sequence for the spin echo data shown in Fig.~\ref{fig:spin_echo} is the typical \(\pi/2-\tau-\pi-\tau-\pi/2\) sequence \cite{childress2006coherent} with modified microwave pulses to address both NV orientations under study (Appendix~\ref{sec:app-exp_details}). The spin echo traces feature an nonuniform sampling with a maximum sampling rate of 3 mHz over the span of the first revival near 50~{\textmu}s. 

We fit each NV center's spin echo trace separately to two fit functions. The first fit function features a strongly coupled \(^{13}\)C and is of the form \cite{childress2006coherent}
\begin{multline}
    f(t)= a_{0} - a_{1}\Big(1-2a_{2}\sin^{2}(\omega_{0}t/2)\sin^{2}(\omega_{1}t/2)\Big)\\
    \times e^{-(t/T_{2})^{b}}\sum_{i=0}^{2}e^{-((t-t_{r}i)/t_{c})^{4}}.\label{eq:spin_echo_osc}
\end{multline}
where \(t=2\tau\) is the total evolution time, the \(a_{i}\) are coefficients, \(T_{2}\) is the coherence time, \(b\) is the exponential stretching factor, \(t_{r}\) is the coherence revival time, and \(t_{c}\) defines the decay time of the revivals. For weak applied magnetic fields, the faster modulation frequency \(2\pi\omega_{0}\) corresponds to \(A_{\text{hfs}}\), the magnitude of the hyperfine interaction between the NV center and a nearby \(^{13}\)C nucleus in the secular approximation \cite{childress2006coherent}. The slower envelope frequency \(2\pi\omega_{1}\) corresponds the \(^{13}\)C Larmor precession frequency, which is effectively enhanced in the presence of the NV center spin. The second fit function is similar to the first, but includes no strongly coupled \(^{13}\)C:
\begin{align}
    f(t)= a_{0} - a_{1}e^{-(t/T_{2})^{b}}\sum_{i=0}^{2}e^{-((t-t_{r}i)/t_{c})^{4}}.\label{eq:spin_echo_no_osc}
\end{align} 

The fitting process for the oscillating fit function (Eq.~\eqref{eq:spin_echo_osc}) consists of three stages. In the first stage the envelope of the trace is extracted by fitting the non-oscillating fit function (Eq.\eqref{eq:spin_echo_no_osc}) to the rolling minimum of the spin echo trace calculated with a window size of 5~{\textmu}s. In the second stage the oscillation parameters \(a_{2}\), \(\omega_{0}\), and \(\omega_{1}\) are determined by brute force optimization. The envelope parameters determined from the first stage are held fixed, with the quartic decay time parameter \(\tau_{c}\) corrected to account for the rolling minimum windows size. The oscillation amplitude \(a_{2}\) is varied from 0 to 1 over 10 steps. The modulation frequency \(\omega_{0}\) is varied from 0 to 2.5 MHz over 2000 steps. The envelope frequency \(\omega_{1}\) is varied from 0 to 1.0 MHz over 2000 steps. Note that the step size in \(\omega_{0}\) is 1.25 kHz, significantly finer than the gaps between the terraces observed in Fig.~\ref{fig:spin_echo}. While we expect that some NV centers should exhibit hyperfine couplings between 5 and 10 MHz, which would be difficult to detect in our ESR data, we limit the modulation frequency \(\omega_{0}\) in the spin echo fits to an upper bound of 2.5 MHz in order to avoid fit biases related to undersampling. The final stage of the fitting process is a gradient-descent optimization over all free parameters simultaneously using the parameters determined from the first and second stages as an initial guess. The fitting process for the non-oscillating fit function consists of just the first and third stages. As described in the main text, we fit to the spin echo traces for 102 NV centers out of the total of 108 NV centers in the group. Of these 102, fits for 6 traces fail to converge, yielding \(\chi^{2}_{\nu}\) above 3 for both fit functions. For the remaining 96 traces, we consider an NV center to have a strongly coupled \(^{13}\)C only if the oscillating fit function yields a \(\chi^{2}_{\nu}\) at least 0.5 lower than the non-oscillating fit function. 

\begin{figure*}[hbtp]
    \includegraphics[width=1\textwidth]{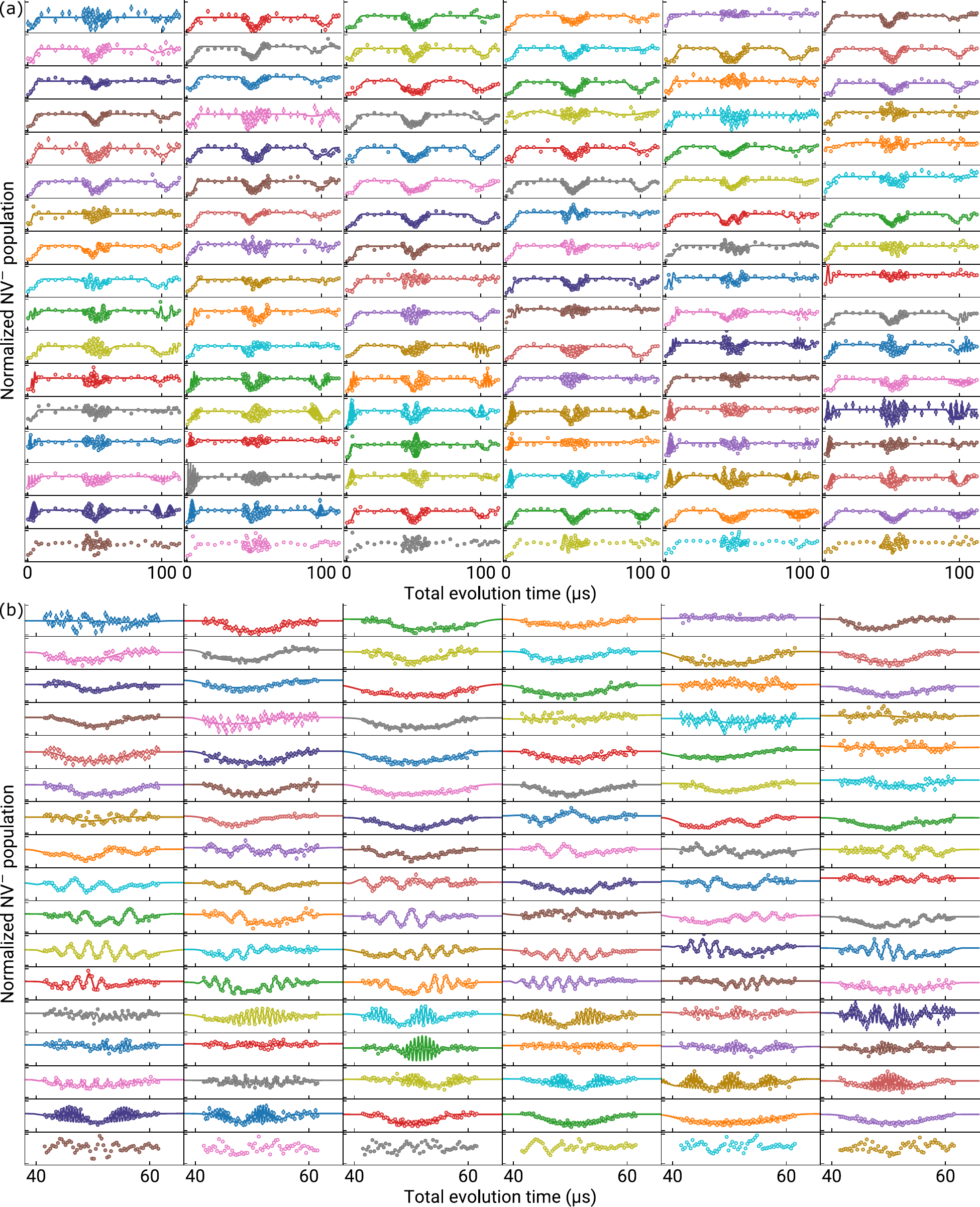}
    \caption{
    Complete set of spin echo traces discussed in Sec.~\ref{sec:parallel} of the main text. Panel (a) shows the full extent of the traces, encompassing two coherence revivals. Panel (b) details the first revival centered at approximately 50~{\textmu}s. In both panels the traces are arranged in order of ascending modulation frequency, as in Fig.~\ref{fig:spin_echo} of the main text. The traces on the bottom row yielded poor fits with \(\chi^{2}_{\nu} > 3\).
    }
    \label{fig:spin_echo-data}
\end{figure*}

The complete set of spin echo traces analyzed in Fig.~\ref{fig:spin_echo} of the main text is shown in Fig.~\ref{fig:spin_echo-data}. All 102 traces are shown, including those for which the fit failed. The traces are arranged in increasing order of \(A_{\text{hfs}}\), as in Fig.~\ref{fig:spin_echo}(b). Traces for the 6 NV centers with significantly broadened or split resonances (bottom row Fig.~\ref{fig:parallel_measurements}) are not shown.

\section{Animated ESR data set}\label{sec:app-animations}

An animated version of the ESR data set shown in Fig.~\ref{fig:parallel_measurements} is available online. The animations show the average background-subtracted EMCCD image as the microwave frequency is varied. The background is the average image recorded for the normalization experiments where the NV centers are prepared in \(m_{s}=0\). The images are downsampled by a factor of 2.

\section{Conditional charge-state initialization}\label{sec:app-conditional_init}

Note that the experiments described in this appendix were conducted using the wide-field version of the apparatus discussed in Appendix~\ref{sec:app-widefield}.

Here we leverage non-destructive parallel charge-state readout and NV center-selective optical pulses in order to demonstrate high-fidelity charge-state initialization of multiple NV centers simultaneously using conditional logic. 

\subsection{Experimental results}\label{subsec:app-conditional_init_exp}

\begin{figure}[!t]
    \includegraphics[width=0.48\textwidth]{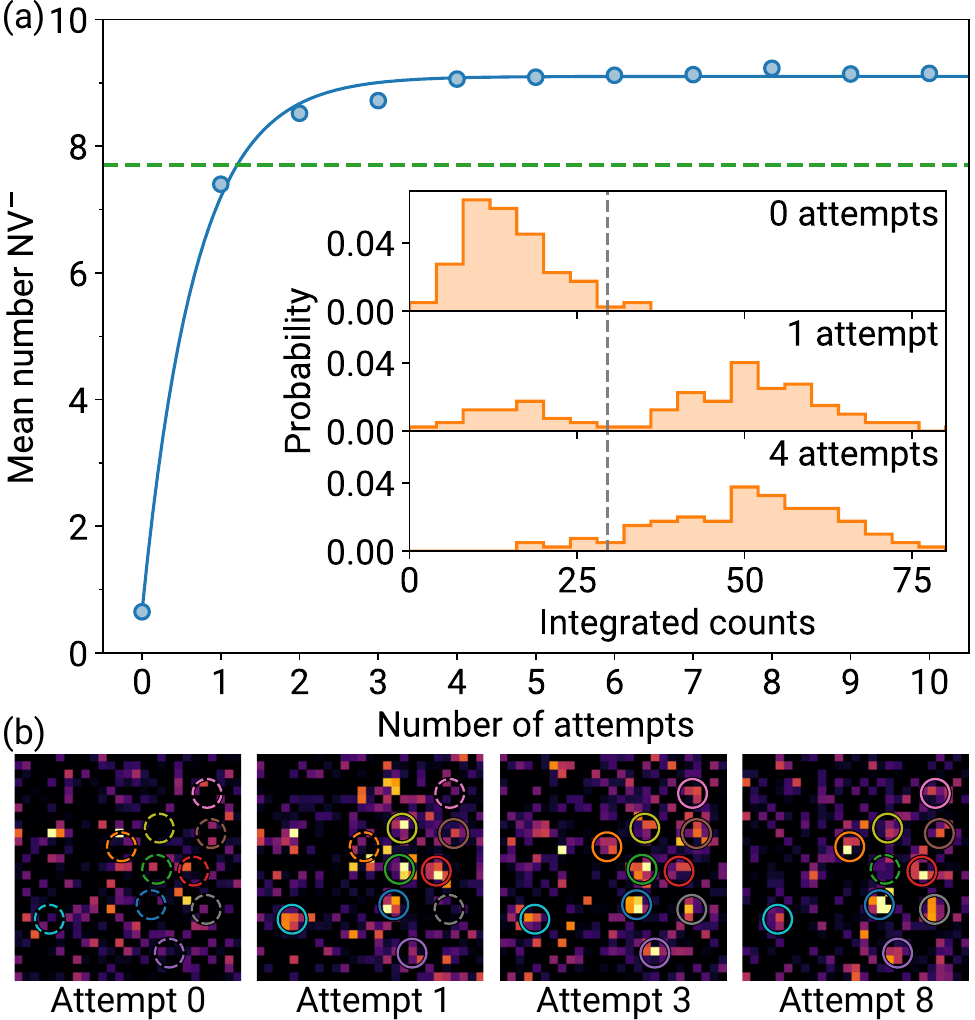}
    \caption{Deterministic initialization of 10 NV centers into NV$^{-}$ under wide-field illumination using conditional logic.
    (a) Scaling of mean number of NV centers in NV$^{-}$ with number of conditional initialization attempts. The NV centers are polarized into NV$^{0}$ at the beginning of each experimental run. 
    After charge-state readout, NV centers found in NV$^{0}$ are re-initialized with 520-nm charge-polarization pulses, and this process of charge-state readout and conditional initialization is repeated. Error bars (one standard error) are smaller than the data points.
    The data is fit to an analytical model (blue curve) described in Appendix~\ref{sec:app-conditional_init}.
    The green dashed line shows the mean number of NV centers found in NV$^{-}$ after unconditional initialization. 
    Inset: Evolution of photon count histograms with number of initialization attempts for a single NV center. The gray dashed line shows the threshold.
    (b) Example single-shot images from the experimental data shown in panel (a). The images are background-subtracted and downsampled such that each NV center occupies roughly four pixels. The background is the average image recorded after polarizing the 10 NV centers into NV$^{0}$. Each NV center used in the experiment is circled with a dashed or solid line according to whether the NV was found in NV$^{0}$ or NV$^{-}$ respectively. 
    }
    \label{fig:conditional_init}
\end{figure}

By repeatedly interrogating the charge states of a group of NV centers and re-initializing only those that are found in NV$^{0}$, the group can be deterministically prepared in NV$^{-}$. This scheme is demonstrated with 10 NV centers imaged under wide-field 589-nm illumination (Appendix~\ref{sec:app-widefield}) in Fig.~\ref{fig:conditional_init}(a), where on average over 9 NV centers are found in NV$^{-}$ after three initialization attempts. In contrast, on average 7.7 NV centers are found in NV$^{-}$ in a separate experiment using unconditional initialization (green dashed line), where before each readout all NV centers are initialized with 520-nm illumination. Example camera single-shot images are shown in fig.~\ref{fig:conditional_init}(b). In the third panel, after three initialization attempts all 10 NV centers were found in NV$^{-}$, representing an instance of ``defect-free'' charge-state initialization of a set of NV centers analogous to the generation of defect-free arrays of neutral atoms via imaging and conditional rearrangement \cite{endres2016atom, barredo2016atom, kim2016situ}. In the fourth image one NV center which had been in NV$^{-}$ appears to have converted to NV$^{0}$. This could be an ionization event induced by the readout light, or a readout error. Unnecessary ionization events during readout could be prevented in the future by targeting for readout only those NV centers which are not yet in NV$^{-}$. 

We note that a 100-ms exposure was used to achieve non-destructive single-shot charge-state readout for the experiments shown in Fig.~\ref{fig:conditional_init}. The experiments described throughout the rest of this work use a 50 ms exposure and a higher 589 nm intensity. While conditional initialization involves a large measurement time overhead, it can still offer improved efficiency over unconditional initialization in certain cases, such as correlated sensing with higher-order joint cumulants \cite{rovny2022nanoscale} or experiments with long interrogation times. Reducing the duration required for non-destructive charge-state readout (e.g., by reducing background using a delta-doped sample) could broaden the range of experiments where conditional initialization is beneficial. For all other experiments described in this work we initialize unconditionally. 

\subsection{Theoretical model}\label{subsec:app-conditional_init_theo}

Here we derive the model used to fit the experimental data shown in Fig.~\ref{fig:conditional_init}(a). The model describes how the number of NV centers measured in NV$^{-}$ scales with the number of initialization attempts. The model includes consideration of ionization during readout, initialization errors, and readout errors. 

As a preliminary, we define readout success probability as the probability of measuring an NV in a specified charge state given that the NV was in that charge state at the beginning of the readout. With \(f_{-}\) and \(f_{0}\) denoting the readout fidelities for charge states NV$^{-}$ and NV$^{0}$ respectively, the following equations describing the relationship between the measured and actual numbers of NV centers in each charge state among a group of \(N\) total NV centers hold:
\begin{gather}
    n = f_{-}n^{\prime} + (1-f_{0})(N-n^{\prime})\label{meas_nv-},\\
    N-n = f_{0}(N-n^{\prime}) + (1-f_{-})n^{\prime}\label{meas_nv0}.
\end{gather}
Here \(n\) (\(n^{\prime}\)) denotes the measured (actual) number of NV centers in NV$^{-}$. The complement \(N-n\) (\(N-n^{\prime}\)) describes the measured (actual) number of NV centers in NV$^{0}$. It is useful to write Eq.~\eqref{meas_nv-} as
\begin{align}
    n^{\prime} = \frac{n - (1-f_{0})N}{(f_{-}+f_{0}-1)}\label{meas_nv-_simp}.
\end{align}

The expected number of NV centers measured in NV$^{-}$ after the \(i\)th initialization attempt is
\begin{align}
    n_{i} &= f_{-}^{2} a n_{i-1}^{\prime} + (1-f_{0})^{2}(N-n_{i-1}^{\prime}) \nonumber\\
    &\quad + f_{-}b(N-n_{i-1}) + (1-f_{0})(1-b)(N-n_{i-1}),\label{cond_init_starting_point}
\end{align}
where \(n_{i}\) (\(n_{i}^{\prime}\)) describes the measured (actual) number of NV centers measured in NV$^{-}$ after the \(i\)th initialization attempt, \(a\) describes the probability that an NV in NV$^{-}$ is not ionized duration readout, and \(b\) describes the probability that an initialization attempt on a given NV center succeeds. Eq.~\eqref{cond_init_starting_point} expresses \(n_{i}\) as a sum of four terms: (1) the number of NV centers that were correctly determined to be in NV$^{-}$ by the previous readout, were not ionized by the previous readout, and are correctly determined to be in NV$^{-}$ by the current readout; (2) the number of NV centers that were incorrectly determined to be in NV$^{-}$ by the previous readout and are again incorrectly determined to be in NV$^{-}$ by the current readout; (3) the number of NV centers that were correctly or incorrectly determined to be in NV$^{0}$ by the previous readout, were then successfully initialized into NV$^{-}$, and are correctly determined to be in NV$^{-}$ by the current readout; and (4) the number of NV centers that were measured in NV$^{0}$ by the previous readout, were then unsuccessfully initialized into NV$^{-}$, and are incorrectly determined to be in NV$^{-}$ by the current readout. 

Eq.~\eqref{cond_init_starting_point} may be written in terms only of the measured values \(n_{i}\) and \(n_{i-1}\) using Eq.~\eqref{meas_nv-_simp}. After simplifying we obtain
\begin{align}
    n_{i} &= c_{1}n_{i-1} + c_{2}N\label{recurrence}
\end{align}
where 
\begin{align}
    c_{1} &= \frac{f_{-}^{2} a - (1-f_{0})^{2}}{(f_{-}+f_{0}-1)} - (f_{-}b + (1-f_{0})(1-b))\\
    c_{2} &= f_{-}b + (1-f_{0})(1-b) + (1-f_{0})^{2} \nonumber\\
    &\quad - \frac{(1-f_{0})(f_{-}^{2} a - (1-f_{0})^{2})}{(f_{-}+f_{0}-1)}.
\end{align}
Solving the recurrence relation of Eq.~\eqref{recurrence}, we arrive at a final model
\begin{align}
    n_{i} &= c_{1}^{i} n_{0} + c_{2}N\frac{1-c_{1}^{i}}{1-c_{1}}.
\end{align}
We note that in the limit of many attempts, the number of NV centers measured in NV\(^{-}\) remains short of the total number of NV centers due to two effects: the probability of ionization during readout, quantified by \(1-a\), and imperfect readout fidelity of the NV\(^{-}\) charge state, quantified by \(f_{-}\). 

We note that in the case of perfect readout fidelity (\(f_{-}=f_{0}=1\)), the coefficients \(c_{1}\) and \(c_{2}\) simplify to 
\begin{gather}
    c_{1} = a - b,\\
    c_{2} = b,
\end{gather}
so that \(c_{1}+c_{2}\) may be interpreted as the probability that readout does not result in ionization, \(a\), with corrections to account for imperfect readout fidelity, and \(c_{2}\) may be interpreted as the initialization success probability, \(b\), with corrections to account for imperfect readout fidelity. The fit parameters extracted from the data shown in Fig.~\ref{fig:conditional_init}(a) are \(n_{0}=0.66(8)\), \(c_{1}=0.225(15)\) and \(c_{2}=0.705(13)\).

\section{Effect of quantum projection noise on contrast of measured correlations}\label{sec:app-qpn}

For the correlation experiments shown in Figs.~\ref{fig:correlations} and \ref{fig:shallow_correlations}, we manually prepared the NV spins in either \(m_{s}=-1\) or \(m_{s}=0\), thereby avoiding quantum projection noise (QPN). As a result, these correlation matrices display the maximum contrast that could be measured between pairs of NV spins with our apparatus given the same experimental parameters (readout fidelities, SCC probabilities, etc.). In real applications, the correlation contrast will be reduced by QPN. 
The effect of QPN in the context of correlated sensing has been thoroughly studied by Rovny et al.~in Ref.~\cite{rovny2022nanoscale}. Here we adapt those results to obtain a simple estimate for how much QPN will reduce the correlation contrast relative to the levels shown in Figs.~\ref{fig:correlations} and \ref{fig:shallow_correlations}. 
For the common case in quantum sensing where an accumulated phase \(\phi\) is mapped onto a population difference by a final \(\pi/2\) pulse before readout (such that the phase is mapped to a polar angle \(\phi + \pi/2\) in the Bloch sphere representation), the measurable correlation between two spins is \cite{rovny2022nanoscale}
\begin{align}
    r = \langle \sin(\phi_{1}) \sin(\phi_{2}) \rangle \label{ideal_correlation}
\end{align}
under ideal conditions of perfect initialization and spin-state readout, and in the absence of uncorrelated noise sources. Here \(\phi_{1}\) and \(\phi_{2}\) are the phases accumulated by the two spins due to a correlated signal. The phases are assumed to be proportional to one another and evenly distributed. The data shown in Figs.~\ref{fig:correlations} and \ref{fig:shallow_correlations} are equivalent to the case where \(\phi_{2}=\pm\phi_{1}\) for any pair of NV centers and \(P(\phi_{1}=\pi/2)=P(\phi_{1}=-\pi/2)=0.5\) for any single NV center, leading to \(r=\pm1\) in the ideal case. For equal phases \(\phi \equiv \phi_{1}=\phi_{2}\) that are normally distributed with variance \(\sigma^{2}_{\phi}\), Eq.~\eqref{ideal_correlation} reduces to \cite{rovny2022nanoscale}
\begin{align}
    r = \frac{1}{2} \big(1-e^{-2\sigma^{2}_{\phi}}\big). \label{gaussian_correlation}
\end{align}
Here \(r\) increases monotonically from 0 at \(\sigma^{2}_{\phi}=0\) to 0.5 for large \(\sigma^{2}_{\phi}\). For large \(\sigma^{2}_{\phi}\) the phases are effectively uniformly distributed on \([-\pi,\pi]\), and therefore each individual spin is completely dephased. In this case the correlation contrast is reduced by a factor of 2 relative to the contrasts shown in Figs.~\ref{fig:correlations} and \ref{fig:shallow_correlations}. We note that non-idealities (e.g., readout errors, initialization errors, uncorrelated noise sources) which will further reduce the correlation contrast can be captured by a prefactor that does not depend on \(r\) itself \cite{rovny2022nanoscale}. As such, the factor by which QPN reduces the contrast in the ideal case will reduce the contrast in the non-ideal case by the same factor for the same distribution of correlated phases.

\begin{figure}[!b]
    \includegraphics[width=0.48\textwidth]{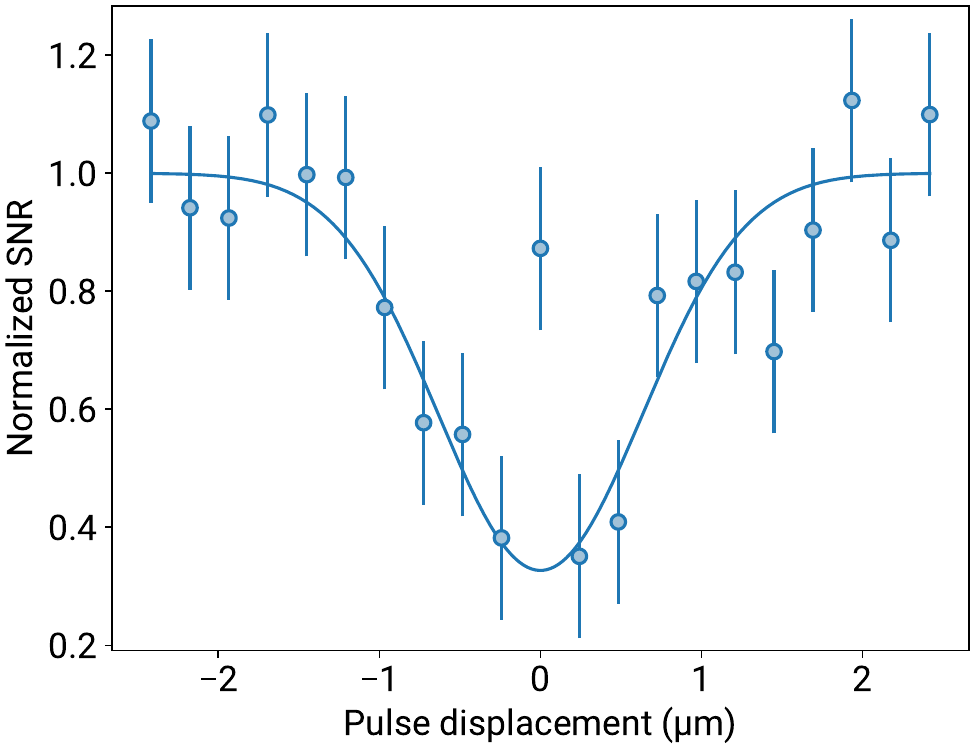}
    \caption{SCC crosstalk characterization.
    The signal-to-noise ratio (SNR) of an NV spin is measured with two SCC pulses applied. The first pulse is displaced relative to the center of the NV center. The second place is not displaced. Solid line is a Gaussian fit to the data excluding the point at zero displacement. The SNR values are normalized to the asymptotic value of the fit. Error bars are one standard error.
    }
    \label{fig:scc_crosstalk}
\end{figure}

\section{Crosstalk characterization}\label{sec:app-crosstalk}

In practice, an optical pulse which is intended to target an individual NV center or a microwave pulse which is intended to target a specific NV orientation may affect other NV centers which are not the target of the pulse.
Here we discuss these crosstalk effects, focusing on the impact to spin measurements conducted using sequences similar to those shown in Figs.~\ref{fig:conditional_init}(a) and \ref{fig:parallel_measurements}(a). 

\subsection{Optical crosstalk} \label{subsec:app-optical_crosstalk}

Note that the experiments described in this appendix sub-section were conducted using the wide-field version of the apparatus discussed in Appendix~\ref{sec:app-widefield}.

Focused 520-nm and 638-nm optical pulses may inadvertently affect both the charge and spin states of NV centers adjacent to a target NV. Charge-state transitions under both 520-nm and 638-nm light are two-photon processes, however, and so depend quadratically on intensity. Therefore, crosstalk affecting the charge state is largely suppressed for optically resolved NV centers. Crosstalk affecting the spin state, on the other hand, can occur at comparably low intensities. 
Because spin polarization is performed globally as a separate step after charge polarization, crosstalk affecting the spin state from 520-nm charge polarization pulses is not relevant to our experiments. Crosstalk from 638-nm SCC pulses can re-polarize the spin states of nearby NV centers that have yet to undergo SCC in the sequence, thereby effectively erasing the information stored in the affected NV's spin state before it is read out. We observed this final optical crosstalk effect in practice as a reduction in the signal-to-noise ratio (SNR) of NV centers with close neighbors. 

SCC crosstalk is quantified by the experiment with a single NV center shown in Fig.~\ref{fig:scc_crosstalk}, where an additional displaced SCC pulse is applied before an SCC pulse which is not displaced in order to simulate the effect of a nearby NV that precedes the target NV in SCC. At large displacements, the SNR approaches the value that would be obtained with no additional SCC pulse. At moderate displacements, the SNR drops as the additional SCC pulse polarizes the target NV center spin. Near zero displacement, the additional pulse performs SCC on the target NV as normal, and the subsequent SCC pulse reduces the population left in NV$^{-}$ regardless of the spin state, thereby similarly reducing the contrast relative to the large-displacement limit. 
A Gaussian fit to the data (excluding the point at zero displacement) reveals a \(1/e^{2}\) crosstalk radius of 1.4(2)~{\textmu}m, significantly larger than the radius of a diffraction-limited 638-nm beam in our setup (about 220 nm). This indicates that NV centers which are optically resolved in EMCCD images are not necessarily sufficiently separated to be unaffected by crosstalk during SCC. The 638-nm beam radius in our apparatus may be larger than the diffraction limit as a result of wavefront errors associated with dichroics, and so the crosstalk radius could potentially be reduced by improving the beam quality. However, because spin polarization is a saturable process, the magnitude of the crosstalk effect does not scale linearly with illumination intensity. For this reason, we predict that even with improvements to the beam quality the crosstalk radius will still be larger than the diffraction-limited beam radius.

\begin{table*}
    \renewcommand{\arraystretch}{1.2}
    \small
    \begin{tabularx}{\textwidth}{|>{\centering\arraybackslash}X >{\centering\arraybackslash}X >{\centering\arraybackslash}X >{\centering\arraybackslash}c >{\centering\arraybackslash}X >{\centering\arraybackslash}X >{\centering\arraybackslash}c |} 
     \hline
     Readout mode & Series/ parallel & \(k\) & \(t_{\text{o,s}}\) (s) & \(t_{\text{i,s}}\) (s) & \(t_{\text{o,p}}\) (s) & \(t_{\text{i,p}}\) (s) \\
     \hline
     conv. & series & 0.03 & \(0.3\times10^{-6}\) & variable & 0 & 0 \\
     conv. & parallel & 0.02 & 0 & 0 & \(0.3\times10^{-6}\) & variable \\
     SCC & series & 0.25 & \(5\times10^{-3}\) & variable & 0 & 0 \\
     SCC & parallel & 0.15 & \(21\times10^{-6}\) & 0 & \(62\times10^{-3}\) & variable \\
     SCC (proj.) & parallel & 0.25 & \(21\times10^{-6}\) & 0 & \(17\times10^{-3}\) & variable \\
     \hline
    \end{tabularx}
    \caption{Variables that determine the time it takes to reach unit SNR for different spin measurement techniques. 
    Symbols: \(k\), single-shot SNR; \(t_{\text{o,s}}\), overhead time for serial operations; \(t_{\text{i,s}}\), integration time for serial measurements; \(t_{\text{o,p}}\), overhead time for parallel operations; \(t_{\text{i,p}}\), integration time for parallel measurements. 
    For conventional readout we assume a 300 ns overhead for readout and polarization. For parallel conventional readout the SNR is reduced to account for the fact that current scientific cameras cannot be gated on \({\sim}100\)~ns timescales and so would need to be continuously exposed during the entire green illumination, rather than only during the window that maximizes SNR. For parallel SCC measurements we assume a dead time of 12 ms per shot for camera readout. 
    The mean value of \(k\) achieved in this work for SLM-based readout was around 0.15 for measurements with both deep NV centers in Sample I and with shallow NV centers in Sample II. With wide-field readout (see Appendix~\ref{sec:app-widefield}), \(k\) varied between 0.2 and 0.3 across 10 deep NV centers within a region of approximately 20~{\textmu}m in diameter in Sample I. We expect that the difference in \(k\) for the two cases stems from imperfections in optical pulse targeting and parameter optimization in the SLM case, which was conducted using an automated script. In contrast, the optical pulses were optimized by hand in the wide-field case. For serial SCC (third row) and the projected scenario for parallel SCC (last row) we take \(k=0.25\). The projected parallel SCC case further assumes that single-shot charge-state readout is achieved  with a 5 ms exposure using a light sheet or delta-doped sample.
    }
    \label{tab:non_correlated_speedup}
\end{table*}

\subsection{Microwave crosstalk} The microwave antenna used in this work \cite{sasaki2016broadband} produces fields that are homogeneous over millimeter scales, and so NV centers over a micron-scale region will experience a uniform field from a microwave pulse. To address two NV orientations we adjust the bias magnetic field to allow for frequency resolution of the \(ms_{s}=0\leftrightarrow m_{s}=\pm1\) transitions of the orientations. We then apply separate microwave pulses at different frequencies for each orientation. Because the pulses are delivered globally, however, NV centers of both orientations will experience the field from a microwave pulse regardless of which orientation the pulse is intended to address, which can result in undesired dynamics. Quantitatively, the amplitude of Rabi oscillations depends on the resonant frequency \(\omega_{0}\), the drive frequency \(\omega_{1}\), and the Rabi frequency \(\Omega\) as
\begin{align}
    A=\frac{\Omega^{2}}{(\omega_{1} - \omega_{0})^{2}+\Omega^{2}}.
\end{align}
The AC Zeeman shift induced by an off-resonant pulse is \cite{ruster2017entanglement}
\begin{align}
    \delta=\frac{1}{4} \Delta m_{s} \Omega^{2}\frac{\omega_{0}}{\omega_{0}^{2} - \omega_{1}^{2}}
\end{align}
for magnetic sublevels which differ in spin projection by \(\Delta m_{s}\). The state with the lower transition frequency relative to \(m_{s}=0\) is used for experiments restricted to two levels. The resonant frequencies for the lower frequency transitions in Sample I are 2.856 GHz for orientation A and 2.800 GHz for orientation B. The Rabi frequency is approximately 10 MHz for both orientations. Therefore we expect population shifts on the order of 3\% and phase shifts of around 71 mrad per \(\pi\) pulse due to microwave crosstalk. The difference in the resonant frequencies in Sample II (Appendix~\ref{sec:app-shallow}) is similar to that of the bulk NV centers, and so similar effects will occur there. 
For sensitive spin experiments which require many microwave pulses (e.g., dynamical decoupling), experiments could be restricted to a single orientation to avoid microwave crosstalk entirely.

\section{Measurement speedup from parallelization}\label{sec:app-speedup}

Here we calculate the speedup enabled by parallelization by estimating the amount of time various measurements would take using different measurement techniques. In particular, we will look the amount of time required to average down a signal to unit signal-to-noise ratio (SNR) for both independent and correlated spin experiments in five different cases: (1) conventional readout with experiments on different NV centers conducted in series; (2) conventional readout with experiments on different NV centers conducted in parallel; (3) SCC readout with experiments on different NV centers conducted in series; (4) SCC readout with experiments on different NV centers conducted in parallel; and (5) projected results for parallel SCC with faster readout and improvements to single-shot SNRs. Conventional readout refers to illuminating an NV center with green light (typically 532 nm) and collecting the resultant spin state-dependent fluorescence. Because green light simultaneously repolarizes the NV spin state, the optimal readout duration for conventional readout is only roughly 200 ns \cite{gupta2016efficient}. Measurements with conventional readout therefore feature low measurement time overhead but also low single-shot SNR.

\begin{figure*}[!t]
    \includegraphics[width=\textwidth]{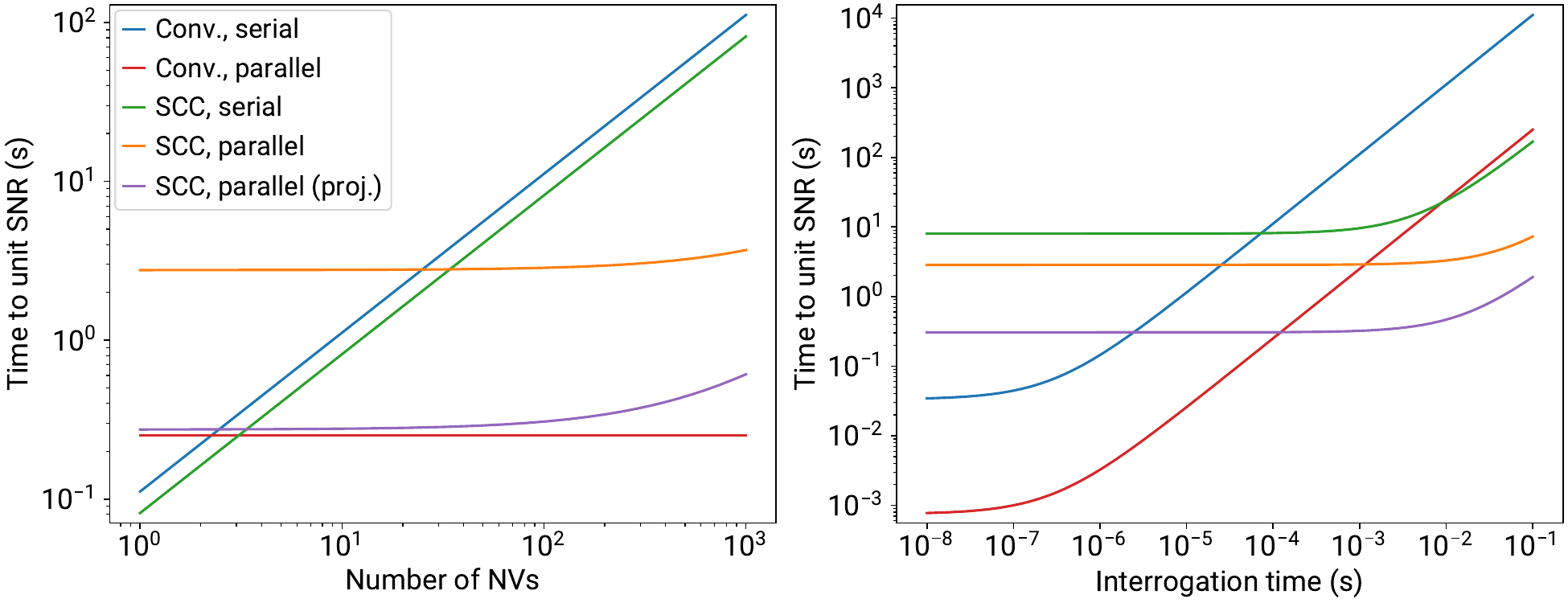}
    \caption{Time to resolve signals for independent spin measurements using different measurement techniques.
    The plots show the amount of time it takes to reach unit SNR according to Eq.~\eqref{non_correlated_speedup} for the techniques described in Table~\ref{tab:non_correlated_speedup}. 
    (a) Required measurement time vs number of NV centers with \(A=1\) and an interrogation time of 100~{\textmu}s, a realistic \(T_{2}\) coherence time for a shallow NV center \cite{wang2016coherence}.
    (b) Required measurement time vs interrogation time with \(A=1\) and \(n=100\) NV centers.
    }
    \label{fig:speedup_independent}
\end{figure*}

\begin{figure*}[!t]
    \includegraphics[width=\textwidth]{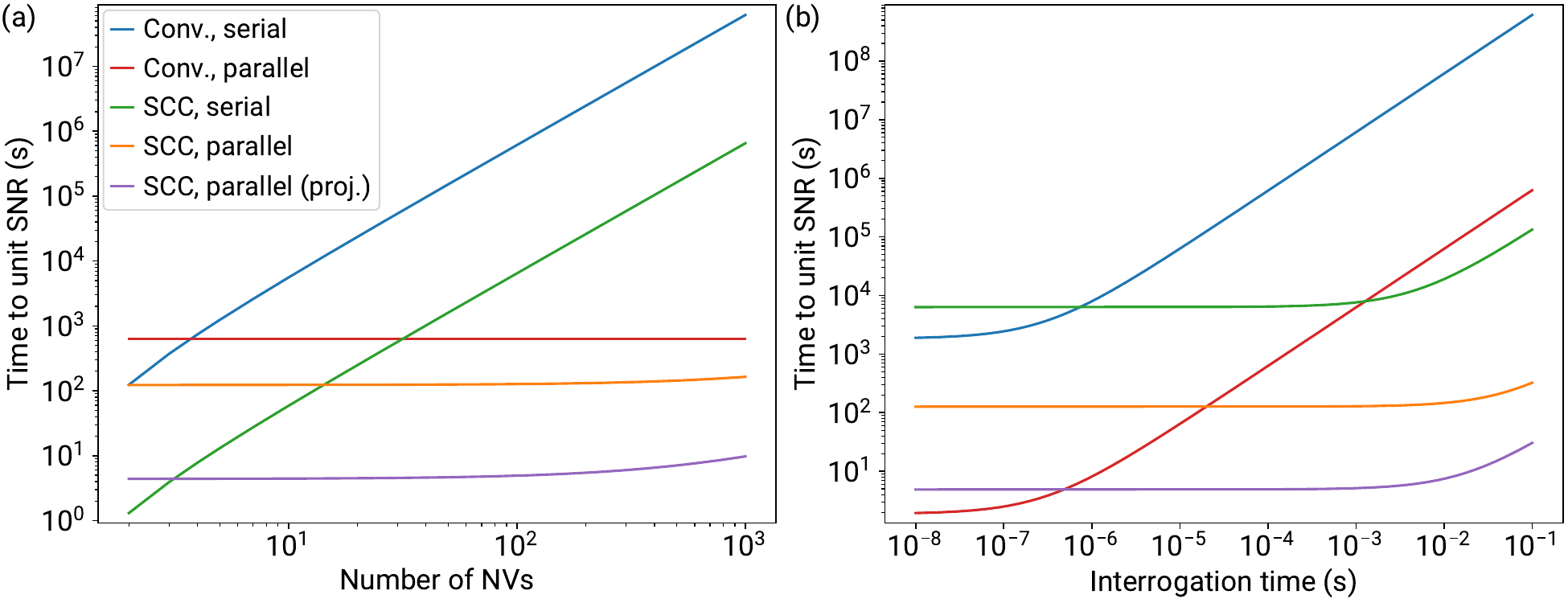}
    \caption{Time to resolve all possible two-point correlators using different measurement techniques.
    The plots show the amount of time it takes to reach unit SNR for the techniques described in Table~\ref{tab:non_correlated_speedup}, according to Eq.~\eqref{correlated_speedup_serial} for the serial techniques and Eq.~\eqref{correlated_speedup_parallel} for the parallel techniques. 
    (a) Required measurement time vs number of NV centers with \(B=1\) and an interrogation time of 100~{\textmu}s.
    (b) Required measurement time vs interrogation time with \(B=1\) and \(n=100\) NV centers.
    }
    \label{fig:speedup_correlated}
\end{figure*}

\subsection{Independent measurements} The time it takes a measurement to reach unit SNR can be expressed generically for both serial and parallel independent spin experiments as 
\begin{align}
    T=A \times (1/k)^{2} \times \left[n\left(t_{\text{o,s}}+t_{\text{i,s}}\right)+t_{\text{o,p}}+t_{\text{i,p}}\right],\label{non_correlated_speedup}
\end{align}
where \(n\) is the number of NV centers, \(t_{\text{o,s}}\) and \(t_{\text{o,p}}\) are overhead times associated with initialization and readout for serial and parallel operations respectively, and \(t_{\text{i,s}}\) and \(t_{\text{i,p}}\) are interrogation times for serial and parallel experiments respectively. The quantity \(k\) is the single-shot SNR defined as the SNR of a single repetition of an experiment in which the NV center is prepared in either the \(m_{s}=0\) or \(m_{s}=+/-1\) spin state, including the effects of state preparation errors \cite{hopper2018spin}. The single-shot SNR is typically inversely proportional to the readout noise discussed in some other works \cite{degen2017quantum, rovny2022nanoscale}. We use the single-shot SNR in our calculations here as it is easily accessed directly in experiment. The quantity \(A\) is a unitless factor that incorporates the target signal strength and additional measurement details. Realistic values for the single-shot SNR \(k\) and time variables are given in Table~\ref{tab:non_correlated_speedup}, and the time required to reach unit SNR according to these parameters is plotted in Fig.~\ref{fig:speedup_independent}. For conventional readout we assume a saturated count rate of 100 kcps, a readout duration of 200 ns, and a spin contrast of 30\%, yielding 0.020 (0.014) photons per readout of the \(m_{s}=0\) (\(m_{s}=\pm 1\)) spin state. We calculate the single-shot SNR \(k\) in this case according to photon counting statistics \cite{hopper2018spin}. For SCC readout we use the single-shot SNR values measured in our experiments where each experimental shot for a single NV center yields one bit based on the inferred charge state. 

\subsection{Correlated measurements} In the case of correlated measurements we examine the required measurement time to reach unit SNR for all \(_{n}C_{2}\) two-point correlators among \(n\) NV centers. The required measurement time can be expressed
\begin{align}
    T=B \times (1/k)^{4} \times \binom{n}{2} \left(t_{\text{o,s}}+t_{\text{i,s}}\right).\label{correlated_speedup_serial}
\end{align}
if the correlators are measured in series using an apparatus like that developed in Ref.~\cite{rovny2022nanoscale} and 
\begin{align}
    T=B \times (1/k)^{4} \times \left[n\left(t_{\text{o,s}}+t_{\text{i,s}}\right)+t_{\text{o,p}}+t_{\text{i,p}}\right].\label{correlated_speedup_parallel}
\end{align}
in the parallel case where all correlators are measured simultaneously. Here, \(B\) is a unitless factor that encodes the magnitude of the correlation due to the target signal, similar to \(A\) in Eq.~\eqref{non_correlated_speedup}. Note the factor of \((1/k)^{4}\) in Eqs.~\eqref{correlated_speedup_serial} and \eqref{correlated_speedup_parallel} \cite{rovny2022nanoscale}, versus the factor of \((1/k)^{2}\) in Eq.~\eqref{non_correlated_speedup} for independent measurements. The results for different measurement techniques are plotted in Fig.~\ref{fig:speedup_correlated}.

\section{Limits to scalability}\label{sec:app-scalability}

In this section we discuss the limits to the scalability of our platform under conditions of improved illumination for charge-state readout by way of a light sheet, delta-doped sample, or similar approach. 

We note that laser power is not a limiting factor for our apparatus. As the 520- and 638-nm beams address just one NV center at a time, only around 5 mW of 520-nm light and 10 mW of the 638-nm is required to reach the sample. The AODs used in this work have diffraction efficiencies of around 40\%. Including an additional 20\% loss from fiber coupling, only around 15 mW is required from the 520-nm laser and 30 mW from the 638-nm laser. A few~{\textmu}W per NV center are needed for 589-nm illumination (Appendix~\ref{subsec:app-pulse_parameters}). The diffraction efficiency of the SLM is approximately 75\%. Including 20\% losses each from fiber coupling and an AOM (for power modulation, see Appendix~\ref{sec:app-exp_details}), the total power required from the laser is around 21~{\textmu}W per NV center, or 2.1 mW for 100 NV centers, if at most 10~{\textmu}W must be delivered to each NV center. 10,000 NV centers would require around 210 mW, which is easily attainable. Faster readout (as in the projected scenario described in Appendix~\ref{sec:app-speedup}) would require higher laser power. For 5-ms readout, roughly 2 watts would be required, which is also attainable.

A key parameter determining the scalability limit is the NV density, \(d\). For the concrete numbers presented in this discussion we assume a diamond sample engineered to contain the maximum practical NV density as limited by the SCC crosstalk radius of 1.4~{\textmu}m. The density in this case is \(d=0.59\) NV centers per square micron, achieved by a hexagonal packing of NV centers separated by 1.4~{\textmu}m. For reference, the NV density for the region of Sample I is \(d=0.16\) NV centers of interest per square micron. While in this work we worked with two specific NV orientations, all four NV orientations could be measured simultaneously using circularly polarized excitation sources. 

\begin{figure}[!t]
    \includegraphics[width=0.48\textwidth]{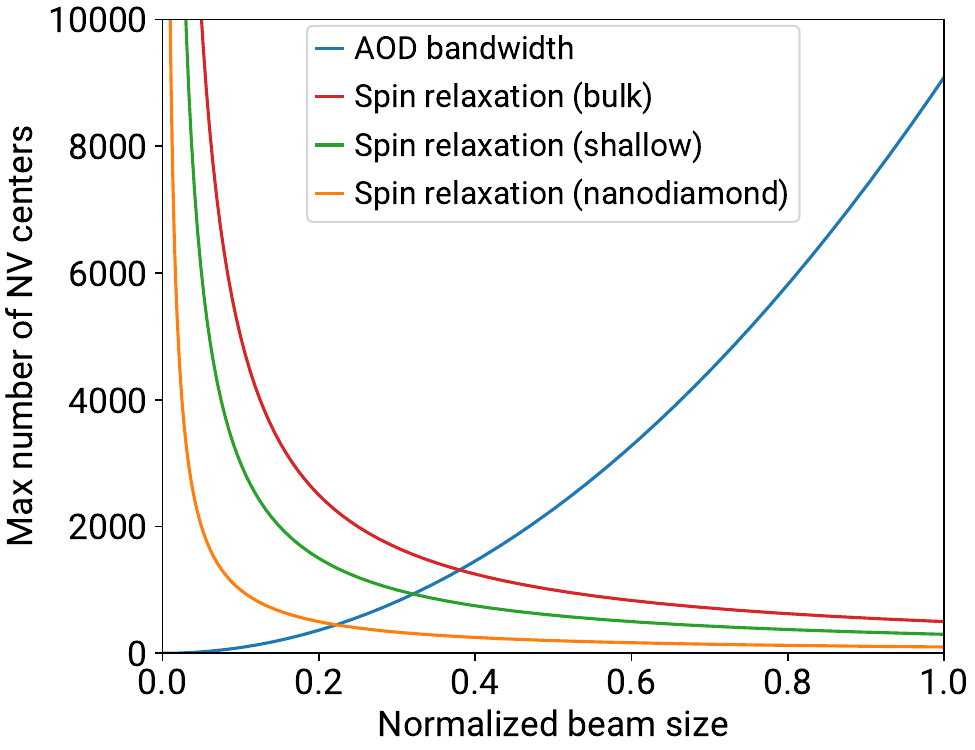}
    \caption{Limits to scalability for different limiting factors.
    The maximum number of NV centers that may be interrogated in parallel depends inversely on the beam size in the spin relaxation-limited case (red, green, and orange curves calculated from Eq.~\eqref{max_n_relaxation}) and quadratically on the beam size in the AOD bandwidth-limited case (blue curve calculated from Eq.~\eqref{max_n_bandwidth} for the 520-nm AOD). The beam size is normalized to the size of the AOD aperture. The intersection points of the curves indicate the optimum beam size and maximum number of NV centers for different SQ relaxation lifetimes. We assume \(1/(3\Omega)=5\)~ms for NV centers deep in bulk diamond, \(1/(3\Omega)=3\)~ms for shallow NV centers, and \(1/(3\Omega)=1\)~ms for NV centers in nanodiamonds. At low temperatures, scalability is limited by the AOD bandwidth to around 9,100 NV centers. 
    }
    \label{fig:scalability_limit}
\end{figure}

\subsection{Spin relaxation and AOD bandwidth} At elevated temperatures, spin relaxation limits the number of NV centers which can be addressed by serial operations before the spin contrast is lost. This effect is not relevant during the polarization step of our experimental sequence, as spin polarization is carried out with a global 589-nm pulse after serial charge polarization. In the SCC section of the sequence, only relaxation on the single-quantum (SQ) \(m_{s}=0 \leftrightarrow m_{s}=\pm 1\) transitions is relevant because SCC does not discriminate between the \(m_{s}=\pm 1\) levels. The time constant for \(1/e\) spin contrast loss associated with SQ relaxation is \(1/(3\Omega)\) where \(\Omega\) is the SQ relaxation rate \cite{myers2017double, gardill2020fast}. At room temperature, this time constant is around 5 ms for NV centers in bulk diamond \cite{cambria2021state}. For near-surface NV centers, accelerated SQ relaxation causes \(1/e\) contrast loss within a few milliseconds for shallow NV centers \cite{myers2017double} and within roughly a single millisecond for NV centers in nanodiamonds \cite{gardill2020fast}. The duration of each SCC operation is dominated by the access time required to effectively update the RF frequencies at the AOD to reposition the 638-nm beam. The AODs used in this work both have nominal access times of \(\tau=10\)~{\textmu}s for beams that fill the AODs' \(7 \times 7\) mm apertures. The maximum number of NV centers that can be addressed by serial SCC before \(1/e\) contrast loss is
\begin{equation}
    \max{n} = \frac{1}{3\Omega \tau} \qquad \text{(relaxation-limited)} \label{max_n_relaxation}
\end{equation}
where \(n\) is the number of NV centers measured in parallel. For an aperture-filling beam Eq.~\eqref{max_n_relaxation} evaluates to 500 NV centers in the bulk context, around 300 NV centers in the shallow context, and around 100 NV centers in the nanodiamond context at room temperature. 

The spin relaxation limit to \(n\) may be increased by reducing the access time of the AOD. The access time is set by the time required for the RF sound wave in the AOD to traverse the beam front, so the access time may be reduced by reducing the beam size. Reducing the beam size at the AOD while maintaining the beam size and angular bandwidth at the back aperture of the objective (by adjusting the lenses in the 4f configuration) would require a corresponding increase in usage of the AOD's RF bandwidth. We only needed 2.4 MHz (2.9 MHz) of RF bandwidth out of the available 40 MHz (45 MHz) for the 638-nm (520-nm) AOD along either of the xy-axes to span the roughly \(8 \times 8\)~{\textmu}m region examined in this work. Therefore, by exchanging the AOD's ``field of view'' for improved access times, we can optimize the maximum of NV centers in the spin relaxation-limited regime. The AOD bandwidth limit to \(n\) can be expressed 
\begin{equation}
    \max{n} = d \times \left(\frac{\partial x}{\partial f}\delta f\right)^{2} \quad \text{(AOD bandwidth-limited)} \label{max_n_bandwidth}
\end{equation}
where \(x\) is the beam position, \(f\) is the RF frequency, and \(\delta f\) is the RF bandwidth. We assume a linear relation between the beam position and the RF frequency over the full RF bandwidth range, and we take \(\partial x / \partial f\) to be the values observed empirically in this work: 3.33~{\textmu}m/MHz for the 638-nm AOD and 2.76~{\textmu}m/MHz for the 520-nm AOD for aperture-filling beams. The 638-nm (520-nm) AOD therefore has an effective field of view of \(0.13\times0.13\)~mm (\(0.12\times0.12\)~mm). Note that we treat the x- and y-channels of the AODs independently here such that the AOD bandwidth-limited sample area is square. Both the access time \(\tau\) and the conversion factor between beam position and RF frequency \(\partial x / \partial f\) scale approximately linearly with the beam size, leading to an optimum beam size that allows the most NV centers to be interrogated in parallel for a given SQ relaxation rate. The result of this tradeoff is illustrated in Fig.~\ref{fig:scalability_limit}. The intersection points between the AOD bandwidth-limited curve and the spin relaxation-limited curves represent the optimum beam size and the maximum number of NVs achievable for each context. From this, the maximum \(n\) is 1100 in the bulk context, 800 in the shallow context, and 400 in the nanodiamond context given the hardware used in this work. We note that for millimeter-scale beams neither the peak diffraction efficiency nor the RF bandwidth is likely to be significantly affected by a reduction in beam size \cite{chang1995acousto}. Small changes in diffraction efficiency that do occur could be compensated for by increasing the RF or laser powers, as the optical powers required for serial operations are relatively low. Another approach to increasing access times is to use an AOD that features a higher speed of sound. Because a higher speed of sound dictates a smaller angular bandwidth for the same RF bandwidth, this approach would similarly require using more of the available RF bandwidth to scan a region of the same area in the sample.

\subsection{Beyond spin relaxation} At cryogenic temperatures the NV spin lifetime can exceed 10 seconds for NV centers deep in high-purity bulk diamond \cite{cambria2023temperature}. In this regime spin relaxation is not a limiting factor (\(\max{n}=10^{6}\)) and so the 638-nm and 520-nm beams should fill the AOD apertures to maximize angular bandwidth, resulting in a limit of around 9,100 NV centers (Fig.~\ref{fig:scalability_limit}). Other potentially limiting factors include the objective field of view, the resolution of the EMCCD camera, the available optical power for parallel operations with 589-nm light, and the potential limits of the specific 589-nm illumination method. The field of view of the objective used in this work is 0.265~mm in diameter, 40\% larger than the diagonal size of the AOD field of view. The EMCCD camera used in this work has a resolution of \(512 \times 512\) pixels. If the diffraction-limited spot for a single NV center (around 500 nm in diameter given our microscope objective) is mapped to a 2 pixel diameter on the camera pixel array, then the total field of view of the camera is \(0.13\times0.13\)~mm, slightly larger than the bandwidth limit for the 520-nm AOD. The 589-nm intensity required in this work was less than 100~W/cm$^{2}$. At this intensity, a 0.3~mm$^{2}$ region could be uniformly illuminated with just 100 mW of optical power. While different illumination methods will require more or less optical power to achieve the same effect, this calculation indicates that the availability of 589-nm optical power is not a significant limit to scalability. Light sheet illumination will not introduce other potential limiting factors to consider. Illumination with an array of focused spots would require a DMD or SLM with a suitably high resolution. We note that SLMs have already been used to generate optical tweezer arrays with over 12,000 sites \cite{manetsch2024tweezer}. We conclude that the upper limit of scalability for our platform is around 9,100 NV centers with hardware similar to that used in this work, and that the limiting factor is AOD bandwidth. 


\bibliography{bibliography.bib}

\end{document}